\documentclass[journal,draftcls,onecolumn,12pt,twoside]{IEEEtranTCOM}
\usepackage{amsmath}
\usepackage{graphicx}
\usepackage{caption}
\usepackage{amsfonts}
\usepackage{url}
\usepackage{array}
\usepackage{bm}
\usepackage{epstopdf}
\usepackage{framed}
\usepackage{tikz}
\usepackage{algorithmic}
\usepackage[]{algorithm}
\usepackage{tabularx}
\usepackage{amsthm}
\usepackage{amssymb}
\usepackage{subcaption}
\usepackage{color} %red, green, blue, yellow, cyan, magenta, black, white
\usepackage[numbers,sort&compress]{natbib}
\usepackage[margin=0.99in]{geometry}
\usepackage{multirow}
\usepackage{alltt}
\renewcommand{\baselinestretch}{1.5}
\usepackage[normalem]{ulem}
\usepackage[utf8]{inputenc}

\usepackage{bigstrut}
\usepackage{booktabs}
\usepackage{alltt}
\usepackage{multirow,booktabs}
\usepackage{multirow} 

\theoremstyle{remark}

\newtheorem{theorem}{Theorem}
\newtheorem{lemma}{Lemma}

\newtheorem{assumption}{}

\definecolor{mycolor1}{rgb}{0.97255,0.97255,0.97255}%
\newcommand{\distas}[1]{\mathbin{\overset{#1}{\kern\z@\sim}}}%

\def \r {{\mathbf r}}
\def \Q {{\mathbf Q}}
\def \EE {{\mathbb E}}
\usepackage{mathtools}

\def \q {{\mathbf q}}
\def \g {{\mathbf g}}

\def \cO {{\mathcal O}}
\def \cX {{\mathcal X}}

\def \cF {{\mathcal F}}

\def \Rn {{\mathbb R}}
\def \lam {{\boldsymbol{\lambda}}}
\def \Lam {{\boldsymbol{\Lambda}}} 
\def \Pib {{\boldsymbol{\Pi}}}
\def \zb {{\boldsymbol{\zeta}}}
\def \xt {\widetilde{\x}}
\def \xh {{\hat{\x}}}

\def \sdF {{\widetilde{\nabla}}} 
\def \dF {{\nabla}}

\def \x {{\mathbf x}}
\def \w {{\mathbf w}}
\def \y {{\mathbf y}}
\def \h {{\mathbf h}}
\def \z {{\mathbf z}}
\def \p {{\mathbf p}}
\def \e {{\mathbf e}}
\def \gb {{\bar{\g}}}
\def \hb {{\bar{\h}}}

\def \L {{\mathbf{L}}}
\def \nb {{\boldsymbol{\nabla}}}
\newcommand{\norm}[1]{\ensuremath{\left\|#1\right\|}}						% norm operator

\usepackage{mathtools}
\newcommand{\genFkl}[1]{\expandafter\newcommand\csname k#1\endcsname{{\mathfrak #1}}}

\renewcommand{\vec}[1]{{\mathbf{#1}}}
\newcommand{\genLatinVec}[1]{\expandafter\newcommand\csname v#1\endcsname{{\vec #1}}}

\newcommand{\mc}[1]{{\mathcal{#1}}}
\newcommand{\mb}[1]{{\mathbb{#1}}}
\newcommand{\genLatinVecU}[1]{\expandafter\newcommand\csname v#1\endcsname{{\vec #1}}}
\def\mydefgreek#1{\expandafter\def\csname v#1\endcsname{\text{\boldmath$\mathbf{\csname #1\endcsname}$}}}
\def\mydefallgreek#1{\ifx\mydefallgreek#1\else\mydefgreek{#1}%
	\lowercase{\mydefgreek{#1}}\expandafter\mydefallgreek\fi}
\mydefallgreek {alpha}{beta}{gamma}{delta}{epsilon}{varepsilon}{zeta}{eta}{theta}{vartheta}{iota}{kappa}{lambda}{mu}{nu}{xi}{omicron}{pi}{rho}{varrho}{sigma}{tau}{upsilon}{phi}{varphi}{chi}{psi}{omega}\mydefallgreek

\def\mydefugreek#1{\expandafter\def\csname v#1\endcsname{\text{\boldmath$\mathbf{\csname #1\endcsname}$}}}
\def\mydefallugreek#1{\ifx\mydefallugreek#1\else\mydefugreek{#1}%
	\lowercase{\mydefugreek{#1}}\expandafter\mydefallugreek\fi}
\mydefallugreek{Gamma}{Delta}{Theta}{Lambda}{Xi}{Sigma}{Upsilon}{Phi}{Psi}{Omega}\mydefallugreek

\newcommand{\vone}{\vec{1}}

\newcommand{\bc}[1]{\left\{{#1}\right\}}
\newcommand{\br}[1]{\left({#1}\right)}
\newcommand{\bs}[1]{\left[{#1}\right]}

\newcommand{\ip}[2]{\left\langle{#1},{#2}\right\rangle}

\newcommand{\E}[1]{\mb{E}\bs{{#1}}}

\newcommand{\argmin}{\mathop{\arg\min}}

\begin{document}
	
	%
	%%%%%%% paper title
	%    \title{Energy Storage Optimization for Real Time Pricing in Smart Grid}
	%    
	\title{Optimal Design of Queuing Systems \\via Compositional Stochastic Programming\footnote{A part of this work has been submitted in Proceedings of the Asilomar Conference on Signals, Systems, and Computers, 2019.}}
%	\author{Srujan Teja Thomdapu and Ketan Rajawat}
	\author{Srujan Teja Thomdapu, ~\IEEEmembership{Student Member,~IEEE} and Ketan Rajawat, ~\IEEEmembership{Member,~IEEE,} \vspace{-1cm}
	\thanks{S. T. Thomdapu, and K. Rajawat are with the Department of Electrical
		Engineering, Indian Institute of Technology Kanpur, Kanpur-208016, India,
		e-mail: ($\{$srujant, ketan$\}$@iitk.ac.in).}}
	
	\maketitle
	\begin{abstract}
		Well-designed queuing systems form the backbone of modern communications, distributed computing, and content delivery architectures. Designs balancing infrastructure costs and user experience indices require tools from teletraffic theory and operations research. A standard approach to designing such systems involves formulating optimization problems that strive to maximize the pertinent utility functions while adhering to quality-of-service and other physical constraints. In many cases, formulating such problems necessitates making simplistic assumptions on arrival and departure processes to keep the problem simple. 
		
		This work puts forth a stochastic optimization framework for designing queuing systems where the exogenous processes may have arbitrary and unknown distributions. We show that many such queuing design problems can generally be formulated as stochastic optimization problems where the objective and constraint are non-linear functions of expectations. The compositional structure	obviates the use of classical stochastic approximation approaches where the stochastic gradients are often required to be unbiased. To this end, a constrained stochastic compositional gradient descent algorithm is proposed that utilizes a tracking step for the expected value functions. The non-asymptotic performance of the proposed algorithm is characterized via its iteration complexity. Numerical tests allow us to validate the theoretical results and demonstrate the efficacy of the proposed algorithm. 
	\end{abstract}
\section{Introduction}
As the Internet traffic continues to grow, meeting service-level agreement (SLA) targets necessitates proactive capacity planning and resource provisioning \cite{buyya2011sla, tan2012provisioning, tan2015adaptive}. Sustaining the users' quality of experience (QoE), for instance, requires the backbone and access networks to be designed with sufficiently high available capacity. Virtual machine resources, such as CPU, memory, and storage, must likewise be provisioned to tackle the demand variability in data centers. On the other hand, economic considerations generally dictate that resources be utilized efficiently and over-provisioning be minimized. Well-designed systems must strive to achieve a balance between resource utilization and excess capacity.

Teletraffic theory, utilizing advances in queuing theory and operations research, has been widely used for the design of communication networks \cite{stidham2009optimal}. Within this framework, the random exogenous factors such as resource demands and channel gains are studied in isolation and statistically modeled a priori (see Chapter 3 of \cite{bertsekas1992data}). Simple models are generally preferred, so as to yield closed-form expressions for various   performance metrics such as waiting times, throughput, delay, waiting costs, provisioning costs, and quality-of-service (QoS). Subsequently, the underlying decision variables, such as the arrival rates or link capacities, are tuned to maximize the performance or minimize the costs, while adhering to the various QoS constraints. The latter step may be accomplished by solving a constrained optimization problem; see \cite{stidham2009optimal} and references therein. 

The classical approach is however not suited to systems where the distribution of the exogenous variables is either unknown or too complicated to allow closed-form expressions. As a simple example, the average ergodic capacity of a single antenna wireless fading channel is not expressible in closed-form \cite{lee2009average}. Consequently, existing approaches for wireless system design either ignore the delay or rely on approximations even for simplistic fading models \cite{chang1995effective,choudhury1996squeezing,wu2003effective, soret2010capacity, champati2018statistical}. More generally, the performance metrics of almost all such queuing systems can be expressed in terms of expectations involving the random variables. This work considers the design of queuing systems from a stochastic optimization perspective. The network design problem is formulated as a constrained optimization problem where the objective and constraint functions involve expectations with respect to these random variables. We consider handling such problems within the stochastic approximation rubric where a few sequentially available independent realizations of the random process are sufficient to yield near-optimal designs. Of particular interest are low-complexity first-order algorithms that are capable of running in real-time. A supervised learning approach is adopted where the parameters are learned during a training phase that is separate from the significantly longer operational phase. The proposed framework is general, and finds applications in communication networks, traffic systems, data centers, and supply chain. 

From a theoretical vantage point, the objective function and constraints  in these design problems usually depend on quantities such as queuing delay and throughput, that are themselves non-linear functions of expectations. The compositional structure of these problems obviates the use of standard stochastic gradient descent (SGD) methods where the gradient estimates are required to be unbiased or at least strongly consistent (see e.g., \cite{benveniste2012adaptive, bertsekas1989parallel, borkar2008stochastic}). The more general case involving non-linear functions of expectations has only been recently considered in \cite{wang2017stochastic}, where the stochastic compositional gradient descent (SCGD) algorithm was proposed. However, compositional problems involving expectation constraints have never been considered in the literature, and the proposed SCGD method for constrained problems is the first such algorithm. The non-asymptotic optimality gap and constraint violation incurred by the proposed algorithm is characterized in terms of the number of samples, and constitutes the key theoretical contribution of the present work. The sample error complexity results developed here are of independent interest and can be applied to any setting involving constrained stochastic optimization problems. 

The rest of the paper is organized as follows. The next subsection briefly reviews the related work on optimization of queuing systems and on algorithms for solving constrained stochastic problems. Sec. \ref{sec-prob} formulates the general constrained problem and discusses several examples from queuing literature. Sec. \ref{sec-scgd} details the proposed algorithm and provides the relevant theoretical guarantees. The simulation results are presented in Sec. \ref{sec-sim} and finally, Sec. \ref{sec-conc} concludes the paper. 

\vspace{-0.5cm}
\subsection{Related Work} 
General compositional stochastic problems, where the goal is to minimize compositions of expected-value functions, were first discussed in \cite{wang2017stochastic} and solved via the SCGD algorithm. The corresponding finite-sum variant of the problem have subsequently been considered in \cite{lian2017finite} and solved via the variance-reduced SCGD. Other variants include the accelerated SCGD \cite{wang2016accelerating}, those utilizing Markov samples \cite{wang2016stochastic}, and generalized versions involving multiple nested compositions \cite{yang2019multilevel}. A functional variant of SCGD has recently been proposed in \cite{koppelcon}. Compositional stochastic problems involving expectation constraints were mentioned in \cite[Sec. 6.10]{ermoliev1988numerical}  but have not been thoroughly studied.

Within the classical stochastic optimization framework, where the compositional form does not arise, problems with stochastic constraints have been considered. Relevant algorithms include the stochastic dual-descent algorithm  \cite{wang2011resource,ribeiro2010ergodic,bedi2018asynchronous,chen2017stochastic} and the stochastic variant of the Arrow–Hurwicz saddle point method \cite{bedi2019asynchronous}. The other problem formulations with infinite time horizon objective and constraint functions are discussed in \cite{wang2016two}. A comprehensive article on constrained stochastic problems of large data can be found in \cite{chen2019learning}. It is remarked that all these classes of algorithms apply to problems with a specific structure, and sample complexity results for the general versions have not been developed. Further, these algorithms cannot be directly applied to problems involving compositional forms. Interestingly, the compositional stochastic optimization problem considered here subsumes the setting with only expectation constraints. 

Optimization approaches for the design of queuing systems have a long history; see \cite{stidham2009optimal} and references therein. Formulated as deterministic optimization problems, these network designs consider a wide variety of objectives and systems. As stated earlier, the existing optimization-based approaches do not suffice when the problem involves random variables with unknown distributions. Optimization of queuing systems where the virtual network functions to be executed in an order are in queue are discussed in \cite{chen2019auto}. The objective function is formulated as a time average processing cost.

The system design problems considered here are static in nature, and the optimal decision variable obtained at the end of the training phase is utilized during the operational phase. Such a setting is in contrast to that of network control, where the goal is to learn optimal policies that depend on the observed environmental state or queue lengths. Utility-optimal control algorithms such as the backpressure algorithm have been extensively studied   \cite{tassiulas1992stability, tassiulas1993dynamic, kumar1995stability, mckeown1999achieving,neely2009optimal, gupta2009low, ying2011combining}. In network control, the optimal policy is updated at every time instant, and the algorithm continues to run forever. However, the focus here is on determining the optimal provisioning or dimensioning parameters that must be set a priori. Accordingly, the performance of the proposed approach is studied in terms of its sample complexity bounds.

\section{Problem Formulation and Examples}\label{sec-prob}
This section formulates the constrained stochastic optimization problem at hand and provides motivating examples within the context of queuing theory. Of particular interest are problems arising in queuing system design, where the distribution of the service times are general and a priori unknown. 
   
%   In this section, we provide mathematical details and few examples of queuing designing models. Classical queuing models can be optimized over parameters such as the arrival rate, service rate, service time distributions, packet length distributions, and channel capacity distributions in wired/wireless networks. The expressions are functions of these parameters. Hence the problems in designing queuing systems can be expressed as, non-linear functions of sample probabilities with functional constraints. A simple example is, the expression for waiting time which is a nonlinear function of service time mean. Closed-Form expressions are readily available in the literature for simple M/M/1 and M/G/1 queues. This paper covers mostly M/G/1 queue examples which have unknown generalized distributions for departure rates and service time. To the best of our knowledge, this framework is not discussed in the existing literature of queuing theory and new from the context of designing queuing systems. The major challenge in this approach is to deal with nonlinear functions of sample probabilities which are also new in the stochastic optimization literature.  
   
%   To this end we will develop a variant of stochastic compositional gradient (SCGD) method and provide mathematical details of it in first subsection. We discuss few examples of M/G/1 queues in wired/wireless networks in later subsection. 
\vspace{-0.5cm}  
\subsection{Problem Formulation}
We begin with describing a general stochastic optimization problem where the formulation involves expectations with respect to a collection of $k$ random variables (collected into the vector $\zb\in\Rn^k$) with unknown distributions. Going beyond the classical stochastic optimization setting where the goal is to minimize the expectation of a function, we consider the more general case where the expectation appears as an argument of an outer non-linear function. In general, such a composite structure makes it difficult to obtain unbiased samples of the objective gradient, rendering the SGD and its variants inapplicable. 

Consider the constrained optimization problem 
   \begin{align} \label{mainProb}
   		\x^\star = \arg\min_{\x \in\cX}~ f(\E{\g(\x,\zb)})\tag{$\mathcal{P}$} \hspace{5mm}\text{s.t. }  \hspace{2mm} \q(\E {\h(\x, \zb)})\leq 0 
   \end{align}  
where the expectation is taken with respect to $\vzeta$. Here, $f:\Rn^m\rightarrow \Rn$, $\g:\Rn^n\times \Rn^k\rightarrow\Rn^m$, $\h:\Rn^n\times \Rn^k\rightarrow\Rn^d$, and $\q:\Rn^d \rightarrow \Rn^J$ are continuous functions. The optimization variable $\x$ belongs to the convex set $\cX \subset \Rn^n$, described by simple constraints that are easy to project onto; examples include box or norm-ball constraints. Specific assumptions regarding convexity or smoothness of various functions will be explicitly stated in Sec \ref{sec-scgd}. In order to avoid trivial scenarios however, we assume that the problem in \eqref{mainProb} is feasible and has a finite solution. 

When the distribution of $\zb$ is not known a priori, the expectations appearing in \eqref{mainProb} cannot be evaluated in closed-form. Motivated by classical stochastic approximation methods, the goal is to solve \eqref{mainProb} in an online fashion using only independent realizations $\zb_1$, $\zb_2$, $\ldots$ that are revealed sequentially. For many problems, such samples yield unbiased (or at least strongly consistent) estimates of the objective gradient, paving the way for the application of the popular SGD algorithm and its variants \cite{ghadimi2013stochastic, chen2018stochastic}. In the present case however, unbiased estimates of the gradient $ \EE[\nb \g(\x,\zb)] \nabla f(\EE \g(\x,\zb))$ cannot be obtained from a single sample $\zb$ due to the presence of the non-linear outer function $f$. Moreover, projected SGD and its variants cannot be used since the constraint also involves an expectation that cannot be evaluated in closed form. 

The set-constrained version of \eqref{mainProb} was first considered in \cite{wang2017stochastic}, where a quasi-gradient approach was proposed. The main idea was to run two iterations: one for performing quasi-gradient steps for estimating $\x^\star$ and another for keeping track of the vector $\E{\g(\x^\star,\vzeta)}$. The approach in \cite{wang2017stochastic} is however still not applicable to \eqref{mainProb} due to the presence of stochastic constraints. 

Stochastic constraints can be dealt with algorithms operating in the dual domain. For instance, dual descent algorithms proposed in \cite{wang2011resource,ribeiro2010ergodic} can handle expectation constraints of a specific form. Likewise, primal-dual or saddle point algorithms have also been proposed for problems with expectation constraints \cite{nedic2009subgradient, koppel2015saddle, bedi2017beyond}. However, none of these approaches can handle non-linear functions of expectations either in the objective or constraint functions. In summary, existing algorithms do not cater to problems of the form in \eqref{mainProb}. Nevertheless, such problems are commonplace in queuing systems where the delay is generally a non-linear function of expectation with unknown distributions.

\subsection{Design of queuing systems }\label{QTE}
% For instance, the queuing delay in an M/G/1 system depends on the second moment of the service time, which in turn may depend on the transmission power allocated to a user. 
%Queuing systems utilizing such an approach must provision for a training phase that precedes the operational phase. More realistically, the environment may be time-varying, where the distribution of the random variable drifts over time, say in a quasi-static manner, and require periodic re-training. Ideally, the training interval must be small, capable of processing the sequence of observations in real-time, and yield the design parameters after a few iterations. 
%Qeuing systems arising in data networks, vehicular traffic flow, and supply chains have been widely studied in the last few decades.	

%	Rapid growth in communication devices over the past several decades demands a congestion efficient data transmission in the network. Queuing theory has been an exciting tool to study system performance parameters and resources in the communication data networks. Queuing theory is the mathematical study first appeared in operations research field and later used in many applications such as design and development of the computer, telecommunication, peer-to-peer \cite{mohajerzadeh2015efficient}, cloud services \cite{tan2015adaptive}, and other networks \cite{nazarathy2012parameter}.

%The problem of optimally designing queuing systems has traditionally been studied from a static vantage point. 

In this section, we describe various queuing design examples that adhere to the  formulation in \eqref{mainProb}.  As a simple example, consider a first-in-first-out (FIFO) M/M/1 queue with a fixed arrival rate $\lambda$ and service rate $\mu$, which is a design variable. As in many packet networks, the system should be designed to balance the utilization or throughput against the delay. Towards this end, consider an objective function that takes the form
\begin{align}
U(\mu) = r\frac{\lambda}{\mu} - h\frac{\lambda/\mu}{\mu-\lambda} \label{simple}
\end{align}
where $r$ and $h$ are positive scaling factors. The first term in \eqref{simple} is proportional to the system throughput while the second term depends on the average delay experienced by a packet while waiting to be served. Maximizing $U(\mu)$ with respect to $\mu$ thus allows the designer to trade off throughput with queuing delay. In the simplest case, it can be seen that the optimal solution is given by $\mu^\star = \lambda + u + \sqrt{u(u+\lambda)}$ where $u = h/r$. In practice, constraints may be imposed so as to restrict $\mu$ to lie within a physically plausible region and may take the form of upper bounds on $\mu$, delay, queue length, and delay variance. Such an optimization framework is general and is applicable to a large class of queuing systems that admit closed-form expressions for utility and delay. See \cite{stidham2009optimal} and \cite{chiang2005geometric} for more examples where deterministic optimization algorithms have been applied to optimize the performance of a queuing systems.

\begin{figure*}
	\setcounter{subfigure}{0}
	\begin{subfigure}{0.5\columnwidth}
		\includegraphics[width=0.9\linewidth, height = 0.5\linewidth]
		{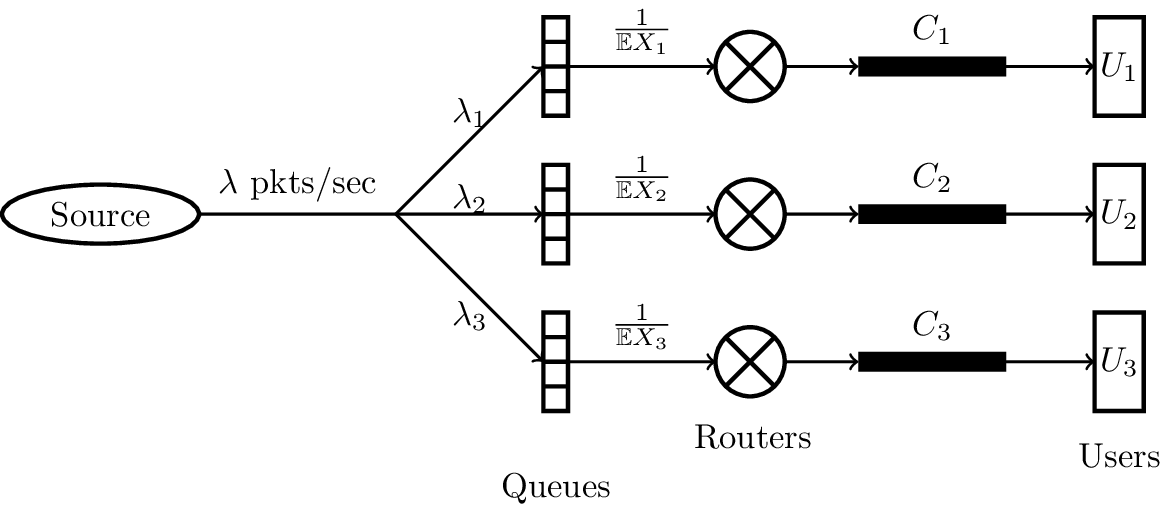}
		\caption{Wired medium}
		\label{mm1SCQ}
	\end{subfigure}
	\begin{subfigure}{0.5\columnwidth}
		\includegraphics[width=0.9\linewidth, height = 0.5\linewidth]
		{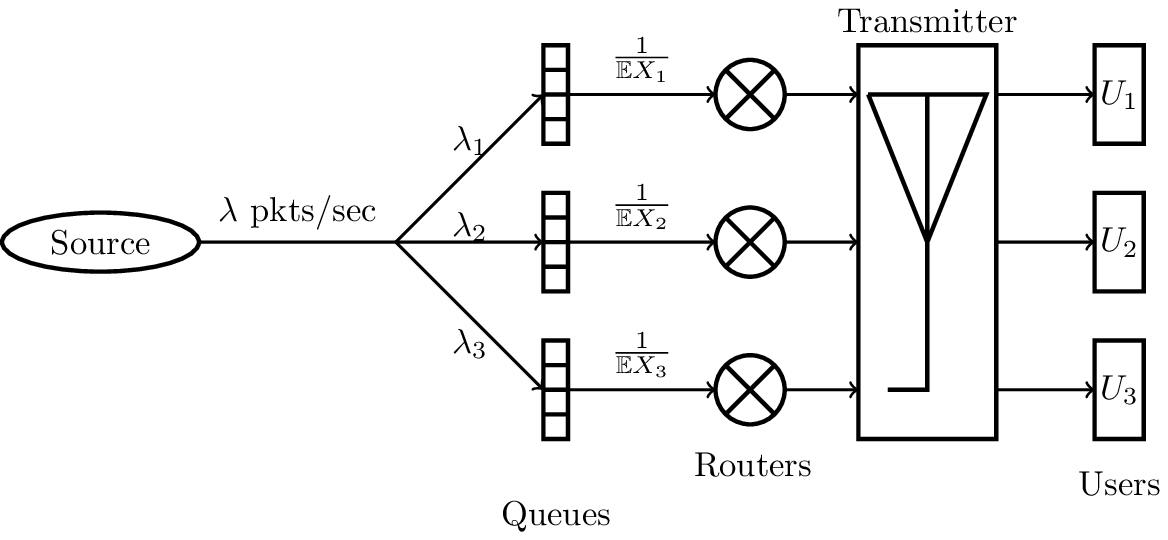}
		\caption{Wireless medium}
		\label{mg1ECQ}
	\end{subfigure}
	\caption{System model for Examples 1-4} 
	\label{sys-model}
	\vspace{-1cm}
\end{figure*}

In this section, several queuing system design examples are discussed and the resulting stochastic problems are shown to be of the form in \eqref{mainProb}. For ease of exposition, we consider the standard utility-minus-penalty objective similar to \eqref{simple} subject to various functional constraints. Other metrics can also be used, provided that they satisfy the assumptions in Sec. \ref{sec-scgd}. Alternatively,  all requirements may be expressed as constraints only, so that \eqref{mainProb} becomes a feasibility problem. 

\subsubsection{Example 1: Parallel M/G/1  queues}\label{mg1s} 
To begin with, consider a set of $N$ parallel $M/G/1$ queues as shown in Fig. \ref{mm1SCQ}. The system consists of a generic source, mimicking a router or an encoder, and capable of supplying at most $\lambda^{\max}$ packets per second on an average. Each packet generated by the source carries a destination address and is accordingly routed to one of the queues. The length of a packet sent to the $i$-th queue is  measured in bits and denoted by the random variable $L_i$ and collected into the random vector $\L$. The lengths are exponentially distributed, but the means $\EE L_i$ are unknown and different for each queue. The difference in mean packet lengths may arise due to the heterogeneity in the traffic stream arriving at each queue. The departures from the $i$-th queue are limited by the link capacity $C_i$ bits per second and the packet service times $X_i := L_i/C_i$.

The optimization variables $\{\lambda_i\}$ are collected into the vector $\lam \in \Rn^N_{++}$ and are required to satisfy box constraints $\lambda_i \in [\lambda^{\min}, \lambda^{\max}]$ that may arise from various physical considerations. As in \eqref{simple}, the goal is to maximize the utility function that  depends on the throughput, and the delay.  While throughput is still given by $\lambda_i\EE L_i$, the queuing delay must be calculated using the Pollaczek-Khinchin (P-K) formula, and the objective function becomes $U(\boldsymbol{\lambda}) = \sum_{i=1}^N \left[\psi_i(\lambda_i\EE L_i) - \varphi\left(\lambda_i\EE L^2_i/\br{2C_i(C_i-\lambda_i\EE L_i)}\right)\right]$, where $\psi_i:\Rn \rightarrow \Rn$ and $\varphi_i:\Rn \rightarrow \Rn$ are pre-specified functions. Examples include linear utility and cost functions, fairness-inducing log utility $\psi(x) = \bar{\psi}_i\log(x)$, and squared delay penalty $\varphi_i(x) = \bar{\varphi}_ix^2$, where $\bar{\psi}_i$ and $\bar{\varphi}_i$ are positive constants.  It can be seen that the objective function depends on a non-linear function of the expectation and can be written in the form required in \eqref{mainProb} as 
\begin{align}\label{fg2}
[\g(\lam, \L)]_i = \begin{cases}
\lambda_i L_i & 1\leq i \leq N \\
\lambda_{i-N}L_{i-N}^2	& N+1\leq i \leq 2N
\end{cases}, \hspace{3mm} f(\y) = \sum_{i=1}^N \left[\varphi\left(\frac{y_{N+i}}{2C_i(C_i-y_i)}\right)-\psi(y_i)\right]
\end{align} 
where $m = 2N$ and $n = k = N$. The waiting delay of a packet should not exceed the maximum tolerable value, and hence the following constraint is imposed:
\begin{align}
\max_{1\leq i \leq N} \frac{\lambda_i\EE L^2_i}{2C_i(C_i-\lambda_i\EE L_i)} \leq D^{\max} \label{maxdelay}
\end{align}
where $D^{\max}$ is the maximum tolerable delay. Comparing with the formulation in \eqref{mainProb}, it can be seen that $\h(\lam,\L) = \g(\lam,\L)$ and $q(\z) = \max_{1\leq i \leq N} z_{i+N}/\br{2C_i(C_i-z_i)}-D^{\max} $ so that $d = 2N$. The overall optimization problem can be written as 
\begin{align*}
\min_{\lam \in \Lam}  ~U(\lam) \hspace{5mm} \text{s.t. }  ~\max_{1\leq i \leq N} \frac{\lambda_i\EE L^2_i}{2C_i(C_i-\lambda_i\EE L_i)} \leq D^{\max},
\end{align*}
where $\Lam:=\{\lam | \lambda^{\min}\leq \lambda_i \leq \lambda^{\max}\forall i=1,\ldots, N, \sum_{i=1}^N\lambda_i \leq \lambda^{lim}\}$.

\subsubsection{Example 2: Parallel M/G/1 queues in wireless networks (ergodic capacity)}\label{m/g/1_ergC}
Consider a system of $N$ parallel M/G/1 queues operating within a common base station (BS) and communicating with $N$ users over $N$ orthogonal channels; see Fig. \ref{mg1ECQ}. The arrivals are still Poisson and the arrival rate vector $\lam$ is an optimization variable. Different from the earlier examples however, the packet lengths are constant and the service time instead depends on the channel capacity and the power allocation. Each queue is allocated with an average power of $p_i$ and the queues adhere to the total power budget $P^{\max}$, that is, $\sum_{i=1}^N p_i \leq P^{\max}.$
The allocated powers are collected into the vector $\p\in\Rn^N_+$. Together, the arrival rates and power allocations belong to the sets $\Lam:= \{\lam | \lambda^{\min}\leq \lambda_i \leq \lambda^{\max}, i=1,\ldots, N, \sum_i \lambda_i \leq \lambda^{\max}\}$ and $\Pib:=\{\p\mid p_i \geq P^{\min}, \sum_i p_i \leq P^{\max}\}$.

The channel gain from the BS to the $i$-th user is denoted by $\zeta_i$, collected in the vector $\zb \in \Rn^N$, and remains constant over the transmission of a single packet. Assuming sufficiently long coherence times, the transmission rate is close to the ergodic capacity of the channel, given by $b_i(p_i, \zeta_i) = B_i\log\left(1+\zeta_ip_i\right)$ $\forall i$,
where $B_i$ is the capacity of the $i$-th channel and is known a priori. If the capacity is measured in packets per second, the service time is the random variable $X_i = 1/b_i(p_i,\zeta_i)$ measured in seconds. The design goal is to determine the optimal values of $\lam$ and $\p$ so as to maximize the utility function. Applying the PK formula for queuing delay, the utility function can be written as
\begin{align}
U(\lam,\p) = \sum_{i=1}^N \left[\psi(\lambda_i) - \varphi_i\left(\frac{\lambda_i \EE\left[\frac{1}{b_i^2(p_i,\zeta_i)}\right]}{2\left(1-\lambda_i\EE\left[\frac{1}{b_i(p_i,\zeta_i)}\right]\right)} \right)\right].\label{mg1EC}
\end{align}
It can be seen that the objective function in \eqref{mg1EC} is of the compositional form in \eqref{mainProb} with 
\begin{align}
[\g(\lam,\p,\zb)]_{i} = \begin{cases}
\lambda_i	& 1\leq i \leq N \\
\tfrac{\lambda_{i-N}}{b_{i-N}(p_{i-N},\zeta_{i-N})} & N+1\leq i \leq 2N \\
\tfrac{\lambda_{i-2N}}{b_{i-2N}^2(p_{i-2N},\zeta_{i-2N})} & 2N+1\leq i \leq 3N \\
\end{cases}, f(\y) = \sum_{i=1}^N\left[\varphi_i\left(\frac{y_{i+2N}}{2(1-y_{i+N})}\right)-\psi(y_i) \right],\label{outerf}
\end{align}
 and $m = 3N$, $n = 2N$, and $k = N$. In order to ensure quality of service to every user, it is also required that the physical layer rate for any user does not drop below a specified level, i.e., $\EE [\min_{i} b_i(p_i,\zeta_i)] > R^{\min}$. In other words, on average, the minimum physical layer rate is not below $R^{\min}$. The constraint is again in the form required in \eqref{mainProb}, with $q(z) = z$, and $h(\lam,\p,\zb) = -\min_{i} b_i(p_i,\zeta_i)$ so that $d = 1$. Unlike in Example 1 however, it can be seen that empirical estimates of the moments of $\zb$ no longer suffice. Instead, the proposed algorithm will entail tracking $\E{\g(\lam^\star,\p^\star,\zb)}$ and $\E{\h(\lam^\star,\p^\star,\zb)}$ explicitly.
 
\subsubsection{Example 3: Parallel M/G/1 queues in wireless networks (outage capacity)}\label{m/g/1_outC}
Ergodic capacity is generally difficult to achieve in practical systems with smaller coherence intervals. We consider the joint design of rate and power for systems where both outage and retransmission events are common. This example builds upon the queuing system described in \cite{ahmed2004throughput}. 

Within the outage-limited setting, a packet in the $i$-th queue is transmitted at a fixed rate $R_i$ regardless of the instantaneous channel conditions. The transmission is deemed successful if $R_i < b_i(p_i,\zeta_i)$ and an outage is declared otherwise. In the case of outage, the server continues to attempt to retransmit a packet until it is successful. Define the indicator function for a successful transmission as  
\begin{align}\label{indout}
r_i(p_i,\zeta_i) = \begin{cases}
1 & R_i< b_i(p_i,\zeta_i) \\
0 & R_i \geq b_i(p_i,\zeta_i).
\end{cases}
\end{align}
Then the service times depend on the number of retransmission attempts, which in turn depends on the probability of successful transmission $\rho_i = \EE[r_i(p_i,\zeta_i)]$. In particular, it can be shown that for this case, the mean waiting time for a packet is given by $w_i(\lambda_i,p_i) = \lambda_i(1+\rho_i)/\br{2R_i(1-\rho_i)(R_i(1-\rho_i)-\lambda_i)}$
%\begin{align}
%w_i(\lambda_i,p_i) = \frac{\lambda_i(1+\rho_i)}{2R_i(1-\rho_i)(R_i(1-\rho_i)-\lambda_i)}
%\end{align}
while the throughput is given by $R_i(1-\rho_i)$. The overall utility function can be written as $U(\lam,\p) = \sum_{i=1}^N\left[\psi(R_i(1-\rho_i)) - \varphi_i\left(w_i(\lambda_i,p_i)\right) \right]$ which is also of the form required in \eqref{mainProb} with 
%\begin{align}
%U(\lam,\p) = \sum_{i=1}^N\left[\psi(R_i(1-\rho_i)) - \varphi_i\left(w_i(\lambda_i,p_i)\right) \right] \label{outut}
%\end{align}
\begin{align}
[\g(\p,\h)]_i &= \begin{cases}
r_i(p_i,\zeta_i) & 1 \leq i \leq N \\
\lambda_{i-N} & N+1\leq i \leq 2N \\
\end{cases} \label{insout}\\
f(\y) &= \sum_{i=1}^N \left[\varphi_i\left(\frac{y_{i+N}(1+y_i)}{2R_i(1-y_i)(R_i(1-y_i)-y_{i+N})}\right) - \psi(R_i(1-y_i)) \right] \label{outout}
\end{align}
Within the context of M/G/1 queues, we briefly discuss some more examples that serve to illustrate the flexibility and generality of the problem considered here. 
\begin{enumerate}
	\item \textbf{Delay variance} captures the jitter and may be an important QoS parameter in streaming and gaming applications. The delay variance can be minimized by adding terms of the form $(\EE[w_i(\lambda_i,p_i)])^2 -\EE[w_i(\lambda_i,p_i)^2]$ in Example 3. Likewise, various moments of delay or throughput may also be incorporated. 
	\item More generally, \textbf{tail probabilities} of the delay or throughput may be considered. For instance in Example 3, where the service time $X_i$ is a random variable, the tail probability can be written as $-\lim_{T\rightarrow\infty}\br{1/T}\mathbb{P}\left( \br{1/T}\sum_{t=1}^T X_{i,t} > x \right)  = \max_{\theta} \left[\theta x - \log \EE[\exp(\theta X_i)]\right]$,
%	\begin{align}
%	-\lim_{T\rightarrow\infty}\frac{1}{T}\mathbb{P}\left( \frac{1}{T}\sum_{t=1}^T X_{i,t} > x \right)  = \max_{\theta} \left[\theta x - \log \EE[\exp(\theta X_i)]\right]
%	\end{align}
	and may therefore be incoporated within the objective function. 
	\item \textbf{Worst-case delay} can be considered by using the term $\max_{1\leq i \leq N} w_i(\lambda_i,p_i)$ in the utility function. Likewise, other non-linear combinations of the delays and throughput may be used. 
	\item \textbf{Expectation constraints} can also be incorporated via penalty methods. For instance, instead of imposing the constraint $\EE[b_i(p_i,\zeta_i)] \geq \epsilon$ in Example \ref{m/g/1_ergC}, it is possible to use a  log-barrier penalty of the form $\log\left(\EE[b_i(p_i,\zeta_i)]-\epsilon\right)$
	can be used in the objective function. Likewise, instead of imposing the constraint of maximum delay $w_i(\lambda_i,p_i)\leq w^{\max}$, a log-barrier penalty of the form $\log\left(w^{\max} - w_i(\lambda_i,p_i)\right)$
	can be used. Log penalties have been widely used in the context of interior point methods. 
	\item \textbf{Probabilistic constraints} can be incorporated much in the same way as in Example \ref{m/g/1_outC}. For instance, a constraint of the form $\mathbb{P}(b_i(p_i,\zeta_i)>R_i)>0.9$ can be converted into an expectation constraint $\EE[r_i(p_i,\zeta_i)] > 0.9$ and dealt with using an appropriate penalty function. 
\end{enumerate}

\subsubsection{Example 4: Parallel G/G/1 queues in wireless networks (effective capacity)}\label{m/g/1_effC}
This example considers the more complicated G/G/1 queue where the expressions for delay cannot be characterized in closed form. The effective capacity, defined as the minimum rate required to guarantee a certain delay (say $W$), is an approximation that allows handling queues with arbitrary arrival and service distributions. Effective capacity has been widely studied for fading channels in systems where a maximum delay of $W$ units may be tolerated \cite{chang1995effective,wu2003effective, soret2010capacity}. 

Let us consider a similar model as in Fig. \ref{mg1ECQ} but with $N$ G/G/1 queues. The packet lengths are constant and the departures depend on the random channel gains and allocated powers $\{p_i\in \Rn_{+}\}_{i=1}^N$. As in Examples 2, 3, the power vector $\p$ is required to lie in the compact set $\Pib$. Different from the earlier examples and for the sake of simplicity, it is assumed that the traffic arrival process at the $i$-th queue has mean $m_i^a$ and variance $(\sigma_i^a)^2$, both of which are known a priori. 

Before formulating the problem, let us briefly review the theory of effective capacity. At any time slot $t$, let $a_i(t)$ denote the instantaneous arrival rate at the $i$-th queue and $c_i(t)$ denote the instantaneous channel capacity of the $i$-th channel.  The corresponding accumulated rates are denoted by $A_i(t) := \sum_{\tau = 1}^{t}a_i(t)$ and $C_i(t) = \sum_{\tau = 1}^{t}c_i(t)$, respectively. The effective capacity of the $i$-th user is given by \cite[Eq. (7)]{wu2003effective}
\begin{align} \label{effC}
	\alpha_i(\theta_i) =  \lim_{t\rightarrow\infty}\frac{1}{tu_i} \log\E{e^{u_iA_i(t)}}|_{u_i = \theta_i}  =   \lim_{t\rightarrow\infty}-\frac{1}{tu_i} \log\E{e^{-u_iC_i(t)}}|_{u_i = \theta_i},
\end{align} 
where $\theta_i$ is the QoS exponent and is the solution to the above equation \cite[Eq. (3)]{soret2010capacity}. 
The steady-state approximation is often used to simplify these expressions. Specifically, as $t\rightarrow \infty$, the application of the central limit theorem allows us to assume the averages $A_i(t)/t$ and $C_i(t)/t$ as Gaussian random variables with finite means and variances. 

In the present case, recall that the instantaneous channel capacity is given by $b_i(p_i,\zeta_i)$ where $\zeta_i$ is the instantaneous channel. Therefore the mean and variance of $C_i(t)/t$ can be approximated as $m^c_i(p_i) = \E{b_i(p_i,\zeta_i)}$, and  $\sigma^c_i(p_i) = \sqrt{\E{b_i^2(p_i,\zeta_i)} - \br{\E{b_i(p_i,\zeta_i)}}^2}$
respectively. As stated earlier, we assume that the means $\{m_i^a\}$ and variances $\{(\sigma_i^a)^2\}$ of the arrival processes are known. Hence \eqref{effC} can be rewritten as \cite[Eq. (6)]{soret2010capacity}
\begin{align}
m^a_i + \frac{u_i}{2}\br{\sigma^a_i}^2|_{u_i = \theta_i} = m^c_i(p_i) - \frac{u_i}{2}\br{\sigma^c_i(p_i)}^2|_{u_i = \theta_i}, 
\end{align}
implying that the QoS quotient $\theta_i$ is given by $\theta_i(p_i) = \br{m^c_i(p_i)-m^a_i}/\br{\br{\sigma^a_i}^2 + \br{\sigma^c_i(p_i)}^2}.$ 
where we have explicated the dependence on the optimization variables $\{p_i\}$. Substituting into \eqref{effC}, we obtain the effective capacity as
\begin{align*}
\alpha_i(p_i) &= \E{b_i(p_i,\zeta_i)} - \frac{\E{b_i(p_i,\zeta_i)}-m_i^a}{\br{\sigma^a_i}^2+ \E{b_i^2(p_i,\zeta_i)} - \br{\E{b_i(p_i,\zeta_i)}}^2} \br{\E{b_i^2(p_i,\zeta_i)} - \br{\E{b_i(p_i,\zeta_i)}}^2}.
\end{align*}
Proceeding along similar lines, the approximation for the probability of delay exceeding the pre-specified value of $W$ time slots is given by $\mb{P}\bc{W_i(t)|_{t\rightarrow\infty} > W} \approx \eta e^{-\theta_i(p_i)\alpha_i(p_i)W}$
where $\eta$ is a normalization constant. So the utility function for this case can be defined as $U(\p) = \sum_{i=1}^N\left[\psi_i\left(\alpha_i(p_i)\right) - \varphi_i\left(e^{-\theta_i(p_i)\alpha_i(p_i)W}\right)\right]$.
The utility is therefore in the compositional form with the inner function
\begin{align*}
	[\g(\p,\zb)]_{i} = \begin{cases}
	b_i(p_i,\zeta_i) & 1\leq i \leq N \\
	b_i^2(p_i,\zeta_i) & N+1\leq i \leq 2N \\
	\end{cases},
\end{align*}
and outer function is given by $f(\y) = \sum_{i=1}^N\bs{  \psi_i\left(\alpha_i(y_i, y_{i+N})\right) + \varphi_i\left(e^{-\theta_i(y_i, y_{i+N})\alpha_i(y_i, y_{i+N})W}\right)}$, where $\theta_i(u,v) = \frac{u-m_i^a}{(\sigma_i^a)^2+v-u^2}$ and $\alpha_i(u,v) = m_i^a + \frac{\theta_i(u,v)}{2}(\sigma_i^a)^2$. In conclusion, it can be seen that the proposed formulation in \eqref{mainProb} is applicable to generalized capacity definitions that also involve non-linear functions of expectations. 

\subsubsection{Example 5: Resource provisioning in cloud services}\label{cloud}
Cloud computing service companies seek to ensure that resources such as power, CPU, and storage are made available to the users as per their demands. Given the uncertainties in the user demands however, the task of distributing the resources to the users is challenging. Ideally, the resources must be provisioned so as to minimize the probability of SLA violation. This example formulates the resource provisioning problem from  \cite{tan2012provisioning,tan2015adaptive} for the case when demand distributions are unknown. 

Consider a service provider that offers $N$ different classes of service and offers a total of $C$ resource units. A customer subscribes to the $i$-th class by paying a service charge of $p_i$ units in return for $r_i$ units of resources. The service provider also incurs a maintenance charge of $\chi$ units per unit time, and per unit resource. An incoming customer may attempt to subscribe to class $i$ of its choice, but may be blocked if the available resources are less than $r_i$ units. The goal is to determine the optimal number of resource units $r_i$ corresponding to each resource class and the total number of resource units to be provisioned.  

Let $\zeta_i$ denote the load for class $i$, which represents the number of customers willing to subscribe to class $i$ if the service provider was not resource-constrained. Then the blocking probability for the $i$-th class is given by \cite[Eq. 1]{tan2015adaptive} $P^B_i(\r,C) = \mb{P}(C-r_i < \sum_{j=1}^N \zeta_j r_j\leq C )/\mb{P}(\sum_{j=1}^N \zeta_jr_j\leq C).$
Roughly, the net revenue obtained by the service provider is given by $R = \sum_i p_i(1-P^B_i(\r,C))$. On the other hand, the maintenance costs incurred by the service provider amount to $ \chi C$ units per unit time. Consequently, the provider's average profit is given by $U(\r,C) = \sum_{i=1}^N p_iN_i(1-P^B_i(\r,C)) - \chi C$, where $N_i$ is the average number of customers actually subscribing to class $i$ per unit time. Finally, the revenue offered for the $i$-th class should be a function of the price $p_i$ for that class. Without loss of generality, let $p_1 < p_2 < \ldots < p_N$. Then, the resource units offered for class $i+1$ should be sufficiently different from that offered for class $i$, and may be restricted to
\begin{align}
r_i + l(p_{i+1}-p_i) \leq r_{i+1} \leq r_i + u(p_{i+1}-p_i).\label{tier}
\end{align}
where $l$ and $u$ are constants. The goal is to maximize the utility $U(\r,C)$ subject to constraints in \eqref{tier}. The problem can be written in the compositional form by defining
\begin{align}
a_i(\r,C,\vzeta) &= \left\{\begin{matrix}
1& \text{if } C-r_i < \zb^T\r \leq C \\ 
0& \text{Otherwise} 
\end{matrix}\right. &
b_i(\r,C,\vzeta) & = \left\{\begin{matrix}
1& \text{if }  \zb^T\r\leq C \\ 
0& \text{Otherwise} 
\end{matrix}\right.
\end{align}
so that the inner function is given by
\begin{align}
[\g(\r,C,\vzeta)]_i  &= \begin{cases}
a_i(\r,C,\vzeta)	& 1\leq i \leq N \\
b_i(\r,C,\vzeta) & N+1\leq i \leq 2N \\
C &  i = 2N+1
\end{cases}
\end{align}
and the outer function is $f(\y) = \sum_{i=1}^Np_iN_i\br{1-\frac{y_i}{y_{N+i}}} - \chi y_{2N+1}$. This example demonstrates the applicability of the compositional form to problems involving probabilistic functions such as $P^B_i(\r,C)$. 

Before concluding, it is remarked that the objective function and constraints in the examples are specifically selected for simplifying the exposition. More general utility functions and constraints are also possible, as long as they adhere to the compositional structure of \eqref{mainProb}.
	
\section{Stochastic Compositional Gradient Descent for Constrained Problems}\label{sec-scgd}
This section details the proposed algorithm for solving \eqref{mainProb} and provides the corresponding convergence rates. For the sake of brevity, we define $F(\x):=f(\E{\g(\x,\zb)})$ and $\Q(\x):=\q(\E{\h(\x,\zb)})$. In the unconstrained case, \eqref{mainProb} can be solved using the SCGD algorithm proposed in \cite{wang2017stochastic} for convex non-smooth or smooth non-convex $F$. This section focuses on the constrained case when both $F$ and $Q$ are convex but possibly non-smooth. 

\vspace{-0.5cm}
\subsection{Assumptions}\label{asm}
We begin with discussing the necessary assumptions on the functions $f$, $\g$, $\q$ and $\h$.

\begin{assumption}\label{adiff}
The outer functions $f$, $\q$ are continuously differentiable and the inner functions $\g$, $\h$ are continuous. Consequently, the sub-gradients of the objective and constraint functions are well-defined, with
	\begin{align}\label{as1}
	\dF F(\x)  := \EE \sdF \g(\x,\zb) \dF f\left(\EE \g(\x,\zb)\right), \hspace{5mm}\dF Q_j(\x) := \EE \sdF \h(\x,\zb) \dF q_j\left(\EE \h(\x,\zb)\right)	
	\end{align}
where $[\sdF \g(\x,\zb)]_{ij} := \partial g_j(\x,\zb)/\partial x_i$ and $[\sdF \h(\x,\zb)]_{ij} := \partial h_j(\x,\zb)/\partial x_i$ with $\partial$ denoting the generalized gradient operation. 
\end{assumption}

\begin{assumption}\label{aslater}
	The problem \eqref{mainProb} is a convex optimization problem and the constraints can be satisfied strictly, i.e., there exists $\xt$ such that $\max_j Q_j(\xt) + \sigma \leq 0$ for some constant $\sigma > 0$. The set $\cX$ is  closed and compact, i.e., $	\sup_{\x,\x' \in \cX} \norm{\x - \x'}^2 \leq D_x < \infty$.
\end{assumption}
\begin{assumption}\label{aiid}
	The random variables $\zb_1, \zb_2,...$ are  independent and identically distributed.  
\end{assumption}
\begin{assumption}\label{ainner}
	The functions $\g$, $\h$ are Lipschitz continuous in expectation and have bounded second order moments, i.e.,
	\begin{align}
	\E{\norm{\sdF \g(\x,\zb)}^2} &\leq C_g, & \E{\norm{\g(\x,\zb) - \EE \g(\x,\zb)}^2} &\leq V_g & \forall &\x \in \cX \label{varG}\\
	\E{\norm{\sdF \h(\x,\zb)}^2} &\leq C_h, & \E{\norm{\h(\x,\zb) - \EE \h(\x,\zb)}^2} &\leq V_h & \forall &\x \in \cX \label{varH}
	\end{align}
\end{assumption}

\begin{assumption}\label{aouter}
	The functions $f$, $q$ are smooth and have bounded gradients, i.e.,
	\begin{align}
	\norm{\dF f(\y)}^2 &\leq C_f, & \norm{\dF f(\y) - \dF f(\bar{\y})} &\leq L_f\norm{\y - \bar{\y}} & \forall &\y,\bar{\y} \in \mb{R}^m \label{gradF}\\
	\norm{\dF \q(\z)}^2 &\leq C_q, &\norm{\dF \q(\z) - \dF \q(\bar{\z})} &\leq L_q\norm{\z - \bar{\z}} & \forall & \z,\bar{\z} \in \mb{R}^d. \label{gradQ}
	\end{align}
\end{assumption}

These assumptions are standard and hold for most problems at hand. In particular, we do not require $\g$ and $\h$ to be smooth or convex. Only the outer functions $f$, $q$ are required to be smooth and only the composite functions $F$ and $\Q$ are required to be convex. As in   \cite{wang2017stochastic}, the matrices $\sdF \g(\x,\zb)$ and $\sdF \h(\x,\zb)$ may consist of gradients, sub-gradients, or other directional gradients satisfying \eqref{as1}. Further, none of the functions are required to be strongly convex. Note that the Lipschitz continous function $\Q$ is defined over a compact domain, and hence is bounded. 

\subsection{Proposed Algorithm}
In order to describe the proposed algorithm, let us also define the smooth and convex function $\ell(\w) = \sum_{j=1}^J \ell_j(w_j+\gamma)$ where $\ell_j$ is defined as
\begin{align} \label{hub}
\ell_j(x) := \begin{cases} \tfrac{1}{2} x^2 & 0 \leq x \leq C_\ell \\
C_\ell x - \frac{C_\ell^2}{2} & x > C_\ell \\
0 & x < 0
\end{cases}
\end{align}
and $\gamma$ and $C_\ell$ are algorithm parameters. It can be seen that $\dF \ell_j(x)=\max\{x,C_\ell\}$ for $x>0$ and zero otherwise. Such a function allows us to define a convex penalty function $L(\x): = \ell(\q(\E{\h(\x,\zb)}))$ that quantifies the constraint violation and allows us to control it through the choice of the parameter $\gamma$. The penalty function defined in \eqref{hub} is quadratic and imposes square penalty on the constraint violation. Intuitively, the update takes the iterate towards the descent direction of optimal function $F$ as well as towards the feasible region $\{\x | \Q(\x) \leq 0\}$.

Recall that the SCGD algorithm for the unconstrained version of \ref{mainProb} entails carrying out updates of the form $\x_{t+1} = \x_t - \alpha_t \hat{\dF} F(\x_t)$
where $\hat{\dF} F(\x_t)$ is an approximation of the actual gradient $\dF F(\x_t)$ and $\alpha_t$ is the step size. To handle the constrained case, we consider the iterations $\x_{t+1} = \x_t - \alpha_t \hat{\dF} F(\x_t) - \delta_t \hat{\dF} L(\x_t)$
where $\delta_t$ is also a step size parameter and $\hat{\dF} L(\x_t)$ approximates the true gradient $\dF L(\x_t)$. Finally, the approximate gradient $ \hat{\dF} L(\x_t)$, and $ \hat{\dF} F(\x_t)$ are calculated by introducing new variables $\z_t$, and $\y_t$ that track the quantities $\E{\h(\x^\star,\zb)}$, and $\E{\g(\x^\star,\zb)}$ are respectively. The step sizes $\alpha_t$, and $\delta_t$ are chosen such that the updates converge to a point that is not only optimal but also feasible. The full updates are summarized in Algorithm \ref{cscgd}. The operation $\Pi_{\cX}$ denotes projection onto the set $\cX$. The computational complexity of the proposed algorithm is only slightly higher than that of the classical stochastic gradient descent method. At each time instant, only gradients of $\g$ and $\h$ are needed in order to carry out the updates. As compared to the SCGD algorithm for unconstrained setting \cite{wang2017stochastic}, the proposed algorithm incurs only a single additional update. Finally, the iteration complexity results specify the precise number of data points needed in order to achieve a specified performance, making the algorithm readily applicable to any real-world application.

\begin{algorithm}
	\caption {Constrained SCGD}
	\begin{algorithmic}[1]
		\STATE\textbf{Input:} $\x_1\in\mathbb{R}^n$, $\y_1\in\mathbb{R}^m$, $\z_1 \in \mb{R}^d$ step sizes $\alpha_t,\beta_t,\delta_t\subset(0,1]$.
		\STATE\textbf{for} $t =1,2,...$
		\STATE\hspace{3mm}Observe the random variable $\zb_t$, and update 
		\begin{align}
		\y_{t+1} &= (1-\beta_t)\y_t + \beta_t \g(\x_t,\zb_t), \hspace{10mm} \z_{t+1} = (1-\beta_t)\z_t + \beta_t \h(\x_t,\zb_t) \label{tr}\\
		\x_{t+1} &= \Pi_{\cX}\bc{\x_t - \alpha_t\sdF\g(\x_t,\zb_t)\dF f(\y_{t+1}) - \delta_t\sdF \h(\x_t,\zb_t)\dF \q(\z_{t+1})\dF\ell(\q(\z_{t+1}))} \label{xup}
		\end{align}
		\STATE\textbf{end}
		\STATE\textbf{Output:} $\xh = \frac{2}{T}\sum_{t = T/2}^T\x_t$.	
	\end{algorithmic}
	\label{cscgd}
\end{algorithm}  

\subsection{Performance Analysis}
This section provides the performance analysis of the proposed algorithm. We begin with describing the surrogate problem
\begin{align} \label{gammaprob}
\x^{\gamma} = \argmin_{\x \in\cX} F(\x), \hspace{5mm} \text{s.t. } \Q(\x) +\gamma \vone\leq 0 
\end{align}
for a given value of $\gamma$. The subsequent analysis utilizes the following key steps (a) bounding the optimality gap $F(\xh)-F(\x^\gamma)$ and the penalty $L(\xh)$; (b) relating the solutions $\x^\star$ and $\x^\gamma$; and (c) obtaining the corresponding bounds on the optimality gap $F(\xh)-F(\x^\star)$ and the constraint violation $\sum_j \max\{Q_j(\xh),0\}$.  

For the first step, we establish the following Lemma, whose proof is provided in Appendix \ref{pfml}.
\begin{lemma}\label{mainLem}
Under Assumptions \textbf{A1}-\textbf{A5}, the iterates generated by Algorithm \ref{cscgd} satsify
	\begin{align}
	\sum_{t = T/2}^T \E{F(\x_t) + \frac{\delta_t}{\alpha_t} L(\x_t) - F(\x^{\gamma})} \leq \cO\left(\frac{1}{\alpha_T} + \sum_{t = T/2}^T\bs{\frac{\delta_t^2}{\alpha_t} + \frac{\beta_t^2}{\alpha_t} + \frac{\delta_t^2}{\alpha_t\beta_t} }\right) \label{jtres}.
	\end{align}
\end{lemma}
The constants within  $\cO(\cdot)$ in \eqref{jtres} depend only on the $D_x,C_q,C_h,C_g,J,C_\ell,L_f,L_q,V_g,$ and $V_h$. The proof of Lemma \ref{mainLem} follows along the lines of the SCGD proof in \cite{wang2017stochastic}, with appropriate modifications incorporated to handle the two variables $\y_t$ and $\z_t$, and the required changes in obtaining the bound. Specifically, the proof requires us to re-define appropriate sequences that allow the telescopic sums to be taken.  

For the next step, we relate the solutions of \eqref{gammaprob} and \eqref{mainProb}. The proof is provided in Appendix \ref{pertproof}.
\begin{lemma}\label{perturb}
	Under Assumption \eqref{aslater} and for $0 < \gamma < \sigma/2$, it holds that $F(\x^{\gamma}) - F(\x^{\star}) \leq 2\sqrt{C_gC_f}D_x\gamma/\sigma$. 
%	\begin{align}
%	F(\x^{\gamma}) - F(\x^{\star}) \leq \frac{2\sqrt{C_gC_f}D_x}{\sigma}\gamma \label{ned}.
%	\end{align}
\end{lemma}

Lemma \ref{perturb} follows from the strong duality of \eqref{gammaprob} and from standard results in duality theory. An implication of Lemma \ref{perturb} is that the gap between the solutions of \eqref{mainProb} and \eqref{gammaprob} cannot be large if $\gamma$ is small. Having stated the preliminary results, we now state the main result of the paper as the following theorem. For the sake of brevity, let us define
	\begin{align}
\omega = 2\frac{D_1}{T\alpha_T} + 2\frac{D_2}{T} \sum_{t = T/2}^T\bs{\frac{\delta_t^2}{\alpha_t} + \frac{\beta_t^2}{\alpha_t} + \frac{\delta_t^2}{\alpha_t\beta_t} } \label{omega}
\end{align}
where $D_1$ and $D_2$ are constants as defined in Appendix \ref{pfml}.

\begin{theorem}\label{thm}
Let the step-sizes be chosen as 
\begin{enumerate}
	\item either diminishing with $\alpha_t = t^{-a}$, $\beta_t = t^{-b}$, $\delta_t = t^{-c}$ for $1\leq t\leq T$; 
	\item or constant with $\alpha_t = T^{-a}$, $\beta_t = T^{-b}$, $\delta_t = T^{-c}$ for all $1\leq t\leq T$,
\end{enumerate}	
with $1 > a \geq c \geq b > 0$. Under \eqref{adiff}-\eqref{aouter} and for $C_\ell$ chosen such that $\max_{\x\in\cX} Q_j(\x) \leq C_\ell-\gamma$, we have the following bounds
\begin{align*}
\E{F(\xh)} - F^\star \leq \omega + \frac{2\sqrt{C_gC_f} D_x}{\sigma} \gamma, \hspace{5mm}
\max_j Q_j(\xh) \leq \sqrt{\frac{ JT^{c-a}\br{\omega + \sqrt{C_fC_g}D_x}}{2^{c-a}}}-\gamma
\end{align*}
where $\omega = \cO\br{T^{a-1} + T^{a-2c} + T^{a-2b} + T^{a+b-2c} }$. 
\end{theorem}

\begin{IEEEproof}
The proof is written in two parts. We first begin with analyzing the optimality gap and subsequently bound the constraint violation. 

%We first begin by finding the relation between $\gamma$ and feasibility gap $\Q(\xh)$ by utilizing result in Lemma \ref{mainLem} followed by, convergence rate is derived with help of Lemma \ref{perturb}. 
Since the right-hand side of \eqref{jtres} is equal to $\omega T/2$ (cf. \eqref{omegaexp}), we have that 
\begin{align}\label{fplusl}
\sum_{t = T/2}^T \E{F(\x_t) + \frac{\delta_t}{\alpha_t} L(\x_t) - F(\x^{\gamma})} \leq \omega \frac{T}{2}.
\end{align}
Since $\E{L(\x_t)} \geq 0$, it follows that $\sum_{t = T/2}^T \E{F(\x_t) - F(\x^{\gamma})}  \leq \omega \frac{T}{2}$. Therefore the optimality gap can be bounded as
\begin{align}
\frac{2}{T}\sum_{t = T/2}^T \E{F(\x_t) - F(\x^{\star})}  \leq \omega  +  \frac{2}{T} \sum_{t = T/2}^T \E{F(\x^{\gamma}) - F(\x^{\star})} \leq \omega + \frac{2\sqrt{C_gC_f} D_x}{\sigma} \gamma. \label{gap}
\end{align}
Since $F$ is convex, the bound in \eqref{gap} implies that $\E{F(\xh)}-F^\star \leq \mathcal{O}(\omega+\gamma)$ where both $\omega$ and $\gamma$ may depend on $T$.

Next observe that since $F$ is Lipschitz with parameter $\sqrt{C_f C_g}$ and the set $\cX$ is compact, it follows that $F(\x_t) - F(\x^{\gamma}) \geq - \sqrt{C_f C_g}\norm{\x_t - \x^{\gamma}} \geq -\sqrt{C_f C_g}D_x$. Substituting into \eqref{fplusl} and rearranging yields
	\begin{align}
\frac{2}{T}\sum_{t = T/2}^T \frac{\delta_t}{\alpha_t}\E{L(\x_t)} \leq \omega + \sqrt{C_f C_g}D_x. \label{feasintm}
\end{align}
Since the sequence $\{\frac{\alpha_t}{\delta_t}\}$ is either diminishing or constant, it holds that $\frac{\alpha_t}{\delta_t} \leq \frac{\alpha_{T/2}}{\delta_{T/2}}$ for all $T/2\leq t\leq T$. Recalling that $\xh = \frac{2}{T}\sum_{t=T/2}^T \x_t$ and that the function $L$ is convex, the left-hand side of \eqref{feasintm} can be lower bounded as
\begin{align}\label{lb}
\frac{2}{T}\sum_{t = T/2}^T \frac{\delta_t}{\alpha_t}\E{L(\x_t)} \geq \frac{\delta_{T/2}}{\alpha_{T/2}} \frac{2}{T}\sum_{t = T/2}^T \E{L(\x_t)} \geq \frac{\delta_{T/2}}{\alpha_{T/2}} \E{L(\xh)}. 
\end{align}	
Recall that $Q$ is bounded and therefore there exists $C_\ell$	such that $\max_j Q_j(\x) \leq C_\ell-\gamma$ for $\gamma \in (0,1)$ or equivalently, $L(\xh) = \sum_{j}[Q_j(\xh)+\gamma]_{+}^2$. Therefore, the lower bound in \eqref{lb} becomes
\begin{align}
\frac{2}{T}\sum_{t = T/2}^T \frac{\delta_t}{\alpha_t}\E{L(\x_t)} \geq \frac{\delta_{T/2}}{\alpha_{T/2}}\sum_{j=1}^J [Q_j(\xh)+\gamma]_{+}^2 
 \geq \frac{\delta_{T/2}}{J\alpha_{T/2}}\left(\sum_{j=1}^J [Q_j(\xh)+\gamma]_{+}\right)^2. \label{lb2}
\end{align}
Substituting \eqref{lb2} into \eqref{feasintm} and taking square root, we obtain
\begin{align*}
\sum_{j=1}^J [Q_j(\xh)+\gamma]_{+} &\leq \sqrt{\frac{J \alpha_{T/2}\br{\omega + \sqrt{C_fC_g}D_x}}{\delta_{T/2}}} \\
\Rightarrow \max_j Q_j(\xh) &\leq \sqrt{\frac{ JT^{c-a}\br{\omega + \sqrt{C_fC_g}D_x}}{2^{c-a}}}-\gamma.
\end{align*}
Finally the required order bound on $\omega$ can be obtained simply by substituting the constant/diminishing step-sizes and ignoring all constant terms. 
\end{IEEEproof}
Theorem \ref{thm} provides a general bound on the optimality gap and the constraint violation that depends on the choice of constants $a$, $b$, $c$, and $\gamma$. For instance, it can be seen that if $\gamma = \sqrt{\frac{ JT^{c-a}\br{\omega + \sqrt{C_fC_g}D_x}}{2^{c-a}}}$
%\begin{align}
%\gamma = \sqrt{\frac{ T^{c-a}\br{\omega + \sqrt{C_fC_g}D_x}}{J2^{c-a}}}
%\end{align}
the proposed algorithm incurs zero constraint violation. 
%Likewise, it is also possible to choose $\gamma = 0$ and incur $\cO(\sqrt{T^{c-a}(\omega+1)})$ constraint violation. 
By carefully choosing the parameters $a$, $b$, and $c$, it is then possible to minimize the bounds on the optimality gap and the constraint violation, as shown in Table  \ref{conrestab}. As expected, the optimality gap decays slowly when stricter requirements are imposed on the constraint violation.  It is remarked that the SCGD algorithm proposed in \cite{wang2017stochastic} for unconstrained problems also incurs an optimality gap of $\mc{O}(T^{-1/4})$, implying that the bounds obtained are tight in the limiting case.  

\begin{table}[]
	\centering
	\caption{Summery of bounds on optimality gap and constraint violation} \label{conrestab}
	\begin{tabular}{|l|l|l|l|}
		\hline
		Choice of constants $a,b,c$ & Optimality gap & Constraint violation & $\gamma$  \\ \hline
		$a = 0.9167$, $b = 0.5$, $c = 0.75$& $\cO\br{T^{-1/12}}$ &0 & $\cO\br{T^{-1/12}}$ \\ \hline
		$a = 0.9167$, $b = 0.5$, $c = 0.75$& $\cO\br{T^{-1/12}}$ &$\cO\br{T^{-1/12}}$ & 0 \\ \hline
		$a = 0.875$, $b = 0.5$, $c = 0.75$& $\cO\br{T^{-1/8}}$ &$\cO\br{T^{-1/16}}$ & 0\\ \hline
		$a = 0.75$, $b = 0.5$, $c = 0.75$& $\cO\br{T^{-1/4}}$ &$\cO\br{1}$ & 0 \\ \hline
	\end{tabular}
\end{table}

 \section{Simulation Results}\label{sec-sim}
 This section provides numerical tests for all examples discussed in Sec. \ref{QTE}. For each example, Algorithm \ref{cscgd} is implemented over a network of $N = 3$ parallel queues for $\psi_i(x) = \bar{\psi}_i\log(x)$, and $\varphi_i(x) = \bar{\varphi}_ix$ for $1\leq i \leq N$. We use the values  $\bar{\psi}_1 = 1$, $\bar{\psi}_2 = 1.5$, $\bar{\psi}_3 = 2$, $\bar{\varphi}_1 = 10$, $\bar{\varphi}_2 = 15$, and $\bar{\varphi}_3 = 20$ in all the examples. The step sizes are chosen as constant with the selection $a = 0.9167$, $b=0.5$, and $c=0.75$. For all cases, the evolution of the optimality gap $F(\xh)-F^\star$ as well as the constraint violation $\max_j Q_j(\xh)$ with iterations is shown.

\subsection{Example 1: Parallel M/G/1 Queues} 
For the problem considered in Example 1, it is clear that outer functions $f$ and $q$ are differentiable (cf. \eqref{fg2} and \eqref{maxdelay}). Likewise, the inner functions $\g$ and $\h$ are continuous, as required in \eqref{adiff}. Further, observe that both the objective function as well as the constraint functions are convex in $\lam$. In its current form however, the gradient of the outer functions $f$ and $q$ is unbounded. To this end, let $L^{\max}$ denote the maximum length of the packet and let $\lambda^{\max}$ be such that $\lambda^{\max} \leq \epsilon C/L^{\max}$ for some given constant $\epsilon \in (0,1)$. Such a requirement ensures that arrival rate is not arbitrarily close to the system capacity.

Given $0<\sigma< D^{\max}$, it can be seen that any $\lambda_i < 2C_i^2(D^{\max}-\sigma)/(\EE L^2_i + 2C_i(D^{\max}-\sigma)\EE L_i)$ is a strictly feasible solution to \eqref{maxdelay}, as required in \eqref{aslater}. Next, \eqref{aiid} holds if the packet lengths are i.i.d. and \eqref{ainner} holds  with $C_g = V_g = C_h = V_h = \sum_{i=1}^N\lambda^{\max}\br{\EE L^2_i + \EE L^4_i}$ where the right-hand side is bounded since $L^{\max} < \infty$. Finally, \eqref{aouter} holds with constants
\begin{align*}
C_f &= \sum_{i=1}^N\bs{\br{\bar{\varphi_i}\frac{1 - \epsilon + \epsilon L_i^{\max}}{4C_i\br{1-\epsilon}^2}}^2 + \br{\frac{\bar{\psi_i}}{\lambda^{\min}}}^2},\\
L_f & = \max_{1\leq i \leq N}\bc{\bar{\varphi_i} \frac{1-\epsilon + \epsilon L_i^{\max}}{C_i^3(1-\epsilon)^3} + \bar{\varphi_i} \frac{\epsilon C_i L^{\max} -1}{2C_i^3(1-\epsilon)^2} + 2\bar{\psi_i} \br{\frac{1}{\lambda^{\min}}}^2 , \bar{\varphi_i}\frac{1}{4C_i^2(1-\epsilon)} }. 
\end{align*}
It can be seen that the constants depend on the specified value of $\epsilon$. In other words, attempt to operate the system very close to the boundary may lead to larger constants and slower rates of convergence. 

\begin{figure*}
	\centering
	\includegraphics[width=0.8\linewidth, height = 0.4\linewidth]{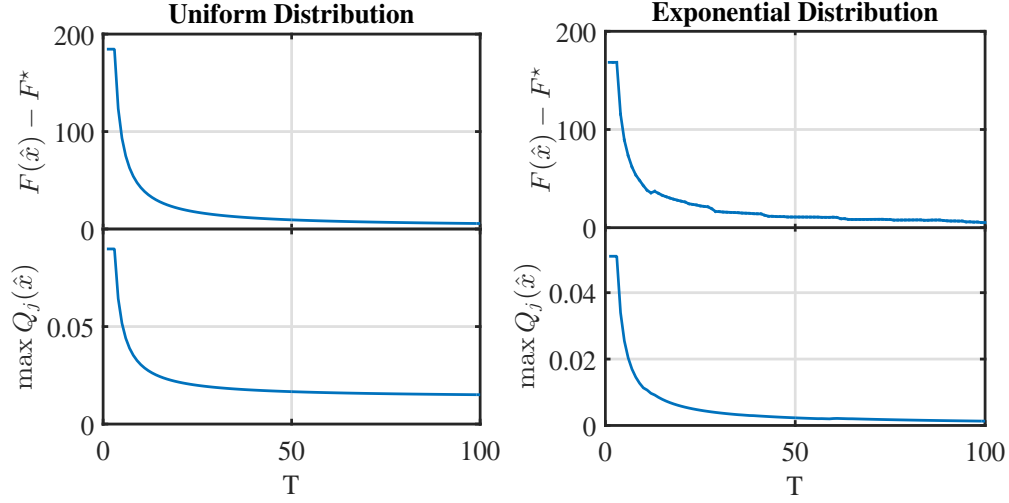}
	\caption{Evolution of the objective function and the constraint violation for Example 1}
	\label{simwired}
\end{figure*}

Fig. \ref{simwired} considers an example system with $D^{\max} = 50 ms$, $C = [100 ~200 ~500]$  kbps, $L^{\max} = [20 ~30~60]$ kb, $\lambda^{\min} = 0.1$, $\lambda = 15$, $\lambda^{\max} = [5~7~9]$, and mean packet lengths $\bar{L} = [15~20~35]$ bits. The simulation is run for different packet length distributions, all truncated to respect the maximum packet length requirements.
It is remarked that Algorithm \ref{cscgd} needs one sample of $L_i$ at each time instant, in addition to the memory required to store the different iterates. As can be seen from Fig. \ref{simwired}, $100$ samples are enough for the algorithm to converge. Note that the problem is simpler if the distribution of packet lengths is known and $F^\star$ can be directly calculated as a function of the first and second moments of $L_i$.  As expected, the optimality gap and the constraint violations decay with iterations. 
 
\subsection{Example 2: Parallel M/G/1 queues in wireless networks (ergodic capacity)}\label{sim_erg}
It can be seen that the inner functions $\g$ and $h$ are continuous as required in \eqref{adiff}. For ease of analysis, let us assume that the support of the random variable $\zeta_i$ is bounded away from zero, i.e., $\zeta_i \geq G$. Coupled with the fact that $p_i \geq P^{\min}$, it follows that $b_i(p_i,\zeta_i) > B_i\log(1+P^{\min}G)$. Therefore, it is always possible to find $p_i \in (P^{\min}, P^{\max})$ such that $\E{\min_i b_i(p_i,\zeta_i)} > R^{\min}-\sigma$ as required in \eqref{aslater}. Further, since $\g$ is locally Lipschitz on compact sets $\Lam$ and $\Pib$, it follows that both $\g (\lam, \p,\zb)$ and $\nabla \g(\lam,\p,\zb)$ are bounded for all $\p \in \Pib$ and $\lam \in \Lam$. Indeed, as required in \eqref{ainner}, we have that 
\begin{align*} 
	C_g = V_g &= N + \sum_{i=1}^N\bigg[\frac{1}{b^2_i(P^{\min},G)}\br{1+\frac{\lambda^{\max}}{\br{1+P^{\min}G}^2\br{b^2_i(P^{\min},G)}}} \\
&\hspace{15mm}+ \frac{1}{b^4_i(P^{\min},G)}\br{1+\frac{2\lambda^{\max}}{\br{1+P^{\min}G}^2\br{b^2_i(P^{\min},G)}}} \bigg],
\end{align*}
$C_h = V_h = \frac{\max_i{B_i}}{\br{1+P^{\min}G}^2}$ where $B_i$ is the capacity of the $i$-th channel. 

Observe that the outer function $f$ as defined in Example 2 is unbounded and has a restricted domain. Instead, we redefine the outer functions as $\tilde{f}$ that approximates $f$ while also satisfying the regularity conditions. Let $\tilde{f}(\y) = \sum_i \varphi_i(\varsigma_i(y_{i+2N},y_{i+N})) - \psi(y_i)$ where
\begin{align*}
\varsigma_i(x,y) = \begin{cases} \tfrac{x}{1-y} & y<1-\epsilon \\
x\tfrac{y-1+2\epsilon}{\epsilon^2} & y \geq 1-\epsilon.
\end{cases}
\end{align*} 
Note that since $b_i(p_i,\zeta_i) > B_i\log(1+P^{\min}G)$, the iterate $\y_t$ takes values in a compact domain, and consequently, it suffices to ensure that $\tilde{f}(\y)$ is locally Lipschitz, implying that $\nabla \tilde{f}(\y)$ is bounded. As required in \eqref{aouter}, we have that 
 	\begin{align*}
C_f &= \sum_{i=1}^N\bs{\frac{\bar{\varphi}_i^2}{4\br{1-\epsilon}^2}\br{1+\frac{1}{b_i^2(P^{\min},G)\br{1-\epsilon}^2}} + \br{\frac{\bar{\psi_i}}{\lambda^{\min}}}^2} \\
L_f &= \max_{1\leq i \leq N}\bc{\frac{\bar{\varphi}_i}{\br{1-\epsilon}^2}\br{\frac{1}{2}+\frac{1}{b_i(P^{\min},G)\br{1-\epsilon}}}  + 2\bar{\psi_i} \br{\frac{1}{\lambda^{\min}}}^2}.
\end{align*}
As $q(z) = z$, it is easy to see that $L_q = 0$, and $C_q = 1$.  Finally, \eqref{aiid} holds if the channel gains are i.i.d.

Verifying the convexity of the redefined objective function is challenging, as the expression for the Hessian is quite complicated. Instead, we resort to a numerical verification by observing that the objective function is smooth in $\lam \in \Lam$ and $\p \in \Pib$. The Hessian for a summand of $\tilde{f}$ is evaluated over a two dimensional grid in $\Lam \times \Pib$ and is found to be always strictly positive definite for the parameters considered in the simulations. Numerical verification is deferred to Appendix \ref{numver}.
%\url{https://tinyurl.com/yxvft4zo}.

 	 \begin{figure*}
		\centering
		\includegraphics[width=0.8\linewidth, height = 0.4\linewidth]
		{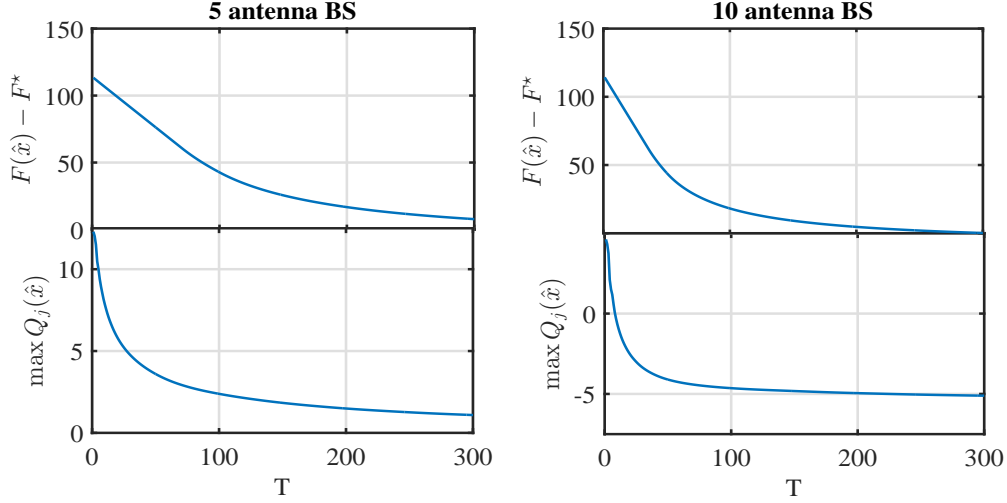}
		\caption{Evolution of the objective function and the constraint violation for Example 2 (ergodic capacity) for 5 antennas and  10 antennas at the BS.}
	\label{simerg}
\end{figure*}
For experiments we consider a setting with $B_i = 10$ kbps for $i = 1, \ldots, N$, $\lambda = 37$, $\epsilon = 0.95$, $G = 0.25$,  $R^{\min} = 35$ kbps, $P^{\min} = 14$ units,  $P^{\max} = 100$ units, $\lambda^{\min} = 0.1$, $\lambda^{\max} = 15$. Rayleigh fading is assumed so that $\zeta_i$ are chi-squared distributed with support truncated to $G$. Two sets of simulations are carried out with $5$, and $10$ antennas at the BS, and the performance is shown in Fig. \ref{simerg}. It can be seen that the optimality gap decays to zero as expected. Interestingly, observe that the constraints are never violated when the number of antennas are large, likely due to the diversity gains available to the system. Different from Example 1 however, $F^\star$ cannot be obtained in closed form but requires brute force search, since the moments of the inner function are themselves functions of the powers. From the Fig. \ref{simerg}, it is evident that for this example, 300 samples are required to obtain a feasible solution that is close to optimum. 
	 
\subsection{Example 3: Parallel M/G/1 queues in wireless networks (outage capacity)}
The objective function in Example 3 is not convex, and therefore the results in Theorem \ref{sec-scgd} do not apply. Nevertheless, the results from \cite{wang2017stochastic} are still applicable under \eqref{adiff}, \eqref{aiid}, \eqref{ainner}, and \eqref{aouter}. First, note that $\Lam$ and $\Pib$ are defined as in Example 2. The channel is also assumed to have a support bounded away from zero, i.e., $\zeta_i \geq G$. Since the domain of $f$ is restricted, it is possible to utilize an approximate version as in Example 2. Alternatively, if the problem parameters satisfy $\epsilon R_i\br{1-r_i(P^{\min},G)} \geq \lambda^{\max}$ for some $\epsilon$, then the outer function's gradient is bounded. In the current form, inner function $g$ (see. \eqref{insout}) is also not continuous as required in \eqref{adiff}. To this end we approximate the indicator function in \eqref{indout} with the sigmoid function $r_i(p_i,\zeta_i) = 1/\br{1+e^{-\eta\br{R_i-b_i(p_i,\zeta_i)}}}$
where $\eta\geq 1$ is a parameter that controls the tightness of the approximation. Outer function $f$ in \eqref{outout} is clearly continuously differentiable (as required in \eqref{adiff}) and \eqref{aiid} holds if the channel gains are i.i.d. Further \eqref{ainner} holds with 
\begin{align}
C_g = V_g = N + \sum_{i=1}^N \frac{1}{2}\br{1+\br{\frac{B_i G}{1+P_i^{\min}G}}^2}.
\end{align}
\begin{figure}
  		\includegraphics[width=0.5\linewidth, height = 0.25\linewidth]
  		{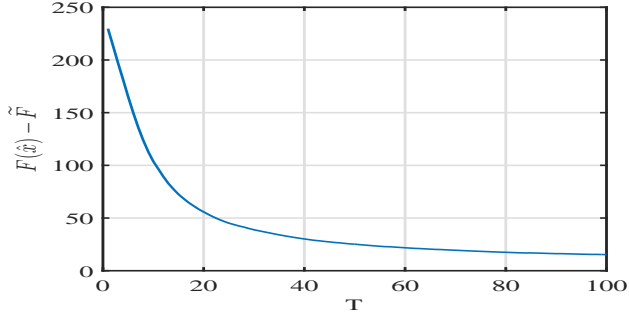}
  		\caption{Evolution of the objective function of Example 3 (outage capacity).}
  		\label{simout}
  	\end{figure}
Since the domain of $(\p,\lam)$ is compact and $f$ is locally Lipschitz, appropriate constants $C_f$ and $L_f$ can likewise be found for \eqref{aouter} as well. 

For experiments we consider a setting with $B_i = 100$ kbps, $\lambda = 45$, $\epsilon = 0.95$, $G = 0.25$,  $R = [30, 35, 40]$ kbps, $P^{\min} = 10$ units,  $P^{\max} = 100$ units, $\lambda^{\min} = 0.1$, and $\lambda_i^{\max} = 25$. Channel gains are exponentially distributed with mean 1 and the base station has a single antenna. The SCGD algorithm \cite{wang2017stochastic} is implemented for this example and the simulations are shown in Fig. \ref{simout}. For this case, the algorithm is only guaranteed to converge to a stationary point. However, in order to also evaluate the quality of the solution, we minimize the objective using the trust-region reflective method as implemented in the fmincon function of MATLAB.  It is remarked that solving the problem using such an algorithm is generally impractical since it requires carrying out simulations in order to calculate the expected gradient and Hessian at every iteration. As evident from Fig. \ref{simout}, it can be seen that the objective function values converge to a local minimum value $\tilde{F}$ obtained from MATLAB, while incurring significantly lower computational cost. 
  
  \subsection{Example 4: Parallel G/G/1 queues in wireless networks (effective capacity)}
\begin{figure}
  		\includegraphics[width=0.5\linewidth,height = 0.25\linewidth]
  		{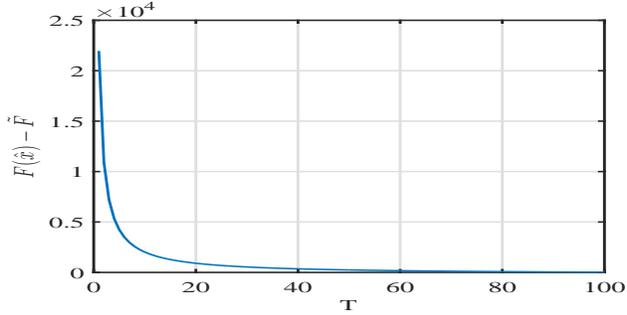}
  		\caption{Evolution of the objective function of Example 4 (effective capacity).}
  		\label{simeff}
  	\end{figure}

   Similar to the previous example, the objective function here is not convex, and the results from \cite{wang2017stochastic} apply under assumptions \eqref{adiff}, \eqref{aiid}, \eqref{ainner}, and \eqref{aouter}. Channel gains at consecutive update intervals are considered to be i.i.d., and therefore \eqref{aiid} holds. As defined in Example 2, 3, $\Pi(\cdot)$ acts as the ball constraint, and because of which the domain of $f$ is restricted. As required in \eqref{aouter}, the outer function's gradient is bounded as long as $\E{b_i(P^{\max},\zeta_i)}$ and $\E{\br{b_i(P^{\max},\zeta_i)}^2}$ are finite. Additionally, outer function $f$ is differentiable as required in \eqref{adiff}. Further \eqref{ainner} holds with
\begin{align}
	C_g = \br{1+4B_i^2} B_i^2 \br{P^{\max}}^2, \hspace{1cm} V_g = \br{1+4B_i^2}B_i^2 \br{P^{\max}}^4.
\end{align}
For experiments, we consider a setting with $B_i = 100$ kbps, $P^{\min} = 0.1$ units, $P^{\max} = 0.9$ units, and $W = 500$ ms. Channel gains are exponentially distributed with means $[0.8,0.9,1]$, and the base station has a single antenna. Simulations for this example are shown in Fig. \ref{simeff}. Since the problem is non convex, the algorithm \cite{wang2017stochastic} is only guaranteed to converge to a stationary point. The local optimum objective $\tilde{F}$ is calculated by using the approximations in \cite{soret2010capacity}. Since the objective function is of composition form, the inner functions $\E{b_i(p_i,\zeta_i)}$, $\E{\br{b_i(p_i,\zeta_i)}^2}$ are approximated to a  deterministic functions of $p_i$, and $\E{\zeta}$,  for Rayleigh channels. In other words, composition form can be avoided if the channel gain $\zeta_i$ is of exponentially distributed. However the approximations are of some complicated functions, and valid only for Rayleigh channels. In contrast, the SCGD algorithm is applicable to any channel with arbitrary distributions. As evident from Fig. \ref{simeff}, the objective function value converges to a local minimum $\tilde{F}$.

\section{Conclusion}\label{sec-conc}
The problem of queuing system design is formulated as a constrained stochastic optimization problem. As the objective and the constraints of the problem are non-linear functions of expectations, classical stochastic approximation methods cannot be used. Instead a constrained variant of the stochastic compositional gradient descent method is proposed where the expected value functions are tracked explicitly. The sample complexity of the proposed algorithm is characterized and it is shown that after $T$ iterations, the average optimality gap and constraint violation are at most $\mc{O}\br{T^{-1/12}}$. Alternative choices of the parameters also yield zero constraint violation but a larger optimality gap of $\mc{O}\br{T^{-1/12}}$ or an $\mc{O}(1)$ constraint violation and the best possible optimality gap of $\mc{O}\br{T^{-1/4}}$. The proposed algorithm is tested over the different queuing design examples and it is shown that the optimality gap and the constraint violation decay as expected. 
   	 
\appendices
\section{Proof of Lemma \ref{mainLem}}\label{pfml}
The overall proof is split into four lemmas. For the sake of brevity, let us define the function $p(\z) := \ell\br{\q(\z)}$. The following preliminary lemma establishes the necessary properties of $p$. 
\begin{lemma}\label{prilL}
The function $p(\z)$ has Lipschitz continuous  and bounded gradients, 
	\begin{align*}
	\norm{\dF p(\z)}^2 \leq JC_qC_{\ell}^2 \hspace{1cm} \norm{\dF p(\z) - \dF p(\bar{\z})} \leq \br{\sqrt{J}L_qC_{\ell} + C_q}\norm{\z - \bar{\z}} \hspace{0.5cm} \forall \z,\bar{\z} \in \mb{R}^d.
	\end{align*} 
\end{lemma}
\begin{IEEEproof}
	From the chain rule, we have that $\dF p(\z) = \dF \q(\z) \dF \ell(\q(\z))$. Using triangle inequality, Assumption \eqref{aouter}, and the definition of $\ell$, we obtain $\norm{\dF p(\z)}^2 \leq JC_qC_\ell^2$ for all $
\z$. In a similar vein, it can be seen that $p$ is smooth since both $\ell$ and $\q$ are smooth functions. In particular, we have that
\begin{align*}
\dF p(\z) - \dF p(\bar{\z}) &= \dF \q(\z)\dF\ell\br{\q(\z)} -  \dF \q(\bar{\z})\dF\ell\br{\q(\bar{\z})}\\
&=  \dF \q(\z)\left(\dF\ell\br{\q(\z)} - \dF\ell\br{\q(\bar{\z})}\right) + \left(\dF \q(\z)-\dF \q(\bar{\z})\right)\dF\ell\br{\q(\bar{\z})}.
\end{align*}
Taking norm, and applying the Cauchy-Schwartz inequality, we obtain
\begin{align*}
\norm{\dF p(\z) - \dF p(\bar{\z})} & \leq \norm{\dF \q(\z)} \norm{\dF\ell\br{\q(\z)} - \dF\ell\br{\q(\bar{\z})}} + \norm{\dF \q(\z)-\dF \q(\bar{\z})}\norm{\dF\ell\br{\q(\bar{\z})}} \\
&\leq \sqrt{C_q}\norm{\q(\z)-\q(\bar{\z})} + L_q\sqrt{J}C_\ell\norm{\z-\bar{\z}}\leq (C_q + \sqrt{J}L_qC_\ell)\norm{\z-\bar{\z}}
\end{align*}
where we have used \eqref{aouter} and the implication that $\q$ is Lipschitz continuous with parameter $\sqrt{C_q}$. 
\end{IEEEproof}

Building upon Lemma \ref{prilL}, we now characterize the consecutive iterate difference and show that it is $\cO(\alpha^2_t+\delta_t^2)$.
\begin{lemma}\label{prilG}
	The updates in \eqref{xup} yields
	\begin{align}
	\E{\norm{\x_{t+1} - \x_t}^2} \leq 2\alpha_t^2 C_fC_g + 2\delta_t^2 J C_{\ell}^2 C_q C_h. 	
	\end{align}	
	for all $t\geq 1$. 
\end{lemma}
\begin{IEEEproof}
	From Algorithm \ref{cscgd}, since $\x_t \in \cX$, the non-expansiveness property of the projection operation implies that
	\begin{align}
	\norm{\x_{t+1} - \x_t}^2 &= \norm{\Pi_{\cX}\bc{\x_t - \alpha_t\sdF\g(\x_t,\zb_t)\dF f(\y_{t+1}) - \delta_t\sdF \h(\x_t,\zb_t)\dF p(\z_{t+1})} - \x_t}^2 \\
	& \leq \norm{\alpha_t\sdF\g(\x_t,\zb_t)\dF f(\y_{t+1}) + \delta_t\sdF \h(\x_t,\zb_t)\dF p(\z_{t+1})}^2\\
	& \leq 2\alpha_t^2\norm{ \sdF\g(\x_t,\zb_t)\dF f(\y_{t+1}) }^2 + 2\delta_t^2 \norm{\sdF \h(\x_t,\zb_t)\dF p(\z_{t+1})}^2\label{ppi}\\
	&\leq 2\alpha_t^2\norm{ \sdF\g(\x_t,\zb_t) }^2 \norm{\dF f(\y_{t+1}) }^2 + 2\delta_t^2  \norm{\sdF \h(\x_t,\zb_t)}^2\norm{\dF p(\z_{t+1})}^2\label{csi}\\
	&\leq 2\alpha_t^2 C_f\norm{ \sdF\g(\x_t,\zb_t) }^2 + 2J\delta_t^2 C_{\ell}^2 C_q \norm{\sdF \h(\x_t,\zb_t)}^2, 
	\end{align}
	where \eqref{ppi} and \eqref{csi} follow from the Peter-Paul and Cauchy-Schwartz inequalities, respectively. The last inequality follows from Lemma \ref{prilL} and \eqref{aouter}. Taking expectations and using \eqref{ainner}, we obtain 	
	\begin{align}
	\E{\norm{\x_{t+1} - \x_t}^2} & \leq 2\alpha_t^2 C_f\E{\norm{ \sdF\g(\x_t,\zb_t) }^2} + 2\delta_t^2 J C_{\ell}^2 C_q \E{\norm{\sdF \h(\x_t,\zb_t)}^2} \\
	&\leq 2\alpha_t^2 C_fC_g + 2\delta_t^2 JC_{\ell}^2 C_q C_h 
	\end{align}
	which is the required result.
\end{IEEEproof}

Next, we establish a key result that legitimizes the tracking steps in \eqref{tr}. To this end, define functions $\gb(\x):=\EE_{\zb}[\g(\x,\zb)]$ and $\hb(\x):=\EE_{\zb}[\h(\x,\zb)]$ for all $\x\in\cX$. Also, let $\cF_t$ be the sigma algebra generated by the random variables $\{\zb_1, \ldots, \zb_{t-1}\}$ so that $\E{\g(\x_t,\zb_t) \mid \cF_t} = \gb(\x_t)$ from \eqref{aiid}.

	\begin{lemma}\label{track}
For the choice $\alpha_t \leq \delta_t$ and $\delta_t \leq \beta_t$, the updates \eqref{tr} in Algorithm \ref{cscgd} yield
\begin{align}
			\hspace{-1cm}\EE\left[\norm{\y_{t+1}-\gb(\x_t)}^2\right] &\leq (1-\beta_t)\E{\norm{\y_t-\gb(\x_{t-1})}^2} + 2V_g\beta_t^2 + 2\frac{\alpha_t^2}{\beta_t} C_fC_g^2 \nonumber \\
			&\hspace{20mm} + 2\frac{\delta_t^2}{\beta_t} JC_{\ell}^2 C_g C_q C_h \leq D_y\label{gtrack}\\
			\hspace{-1cm}\EE\left[\norm{\z_{t+1}-\hb(\x_t)}^2\right] &\leq (1-\beta_t)\E{\norm{\z_t-\hb(\x_{t-1})}^2} + 2V_h\beta_t^2 + 2\frac{\alpha_t^2}{\beta_t}C_fC_gC_h \nonumber \\
			&\hspace{20mm} +  2\frac{\delta_t^2}{\beta_t}JC_{\ell}^2 C_q C_h^2 \leq D_z \label{htrack}
			\end{align}
where $D_y$ and $D_z$ are initialization dependent constants. 
	\end{lemma}
\begin{IEEEproof}
We provide the proof of \eqref{gtrack} while observing that \eqref{htrack} follows along similar lines. Let $\e_t:=(1-\beta_t)\left(\gb(\x_t)-\gb(\x_{t-1})\right)$ be the difference between evaluations of $\gb$ at consecutive iterates, scaled by $(1-\beta_t)$. From the Peter-Paul inequality, we have that
\begin{align}
\norm{\y_{t+1} -\gb(\x_t) }^2 \leq (1+\beta_t)\norm{\y_{t+1} -\gb(\x_t) + \e_t}^2+(1+\frac{1}{\beta_t})\|\e_t\|^2.\label{pep}
\end{align}
Since $\gb$ is Lipschitz (cf. \eqref{ainner}), the second term of \eqref{pep} can be bounded as
	\begin{align}
	\norm{\e_t} \leq (1-\beta_t)\sqrt{C_g}\norm{ \x_t-\x_{t-1} }.
	\end{align}
For the first term of \eqref{pep}, we have that
	\begin{align}
	\y_{t+1} -\gb(\x_t) + \e_t = (1-\beta_t)\left(\y_t-\gb(\x_{t-1})\right) + \beta_t\left(\g(\x_t,\zb_t) - \gb(\x_t)\right).
	\end{align}
	Taking squared norm, applying conditional expectation given $\cF_t$, and using the definition of $\gb$, we obtain
	\begin{align}
	\E{\norm{\y_{t+1} -\gb(\x_t) + \e_t}^2\mid \cF_t}&= \E{\norm{\left(1-\beta_t\right)\left(\y_t-\gb(\x_{t-1})\right) + \beta_t\left(\g(\x_t,\zb_t) - \gb(\x_t)\right)}^2 \mid \cF_t} \nonumber\\
	& = (1-\beta_t)^2\norm{\y_t-\gb(\x_{t-1})}^2 + \beta_t^2\EE\left[\norm{\g(\x_t,\zb_t) - \gb(\x_t)}^2\rvert\cF_t\right]	\nonumber\\
	&\hspace{5mm}+ 2(1-\beta_t)\beta_t\left(\y_t-\gb(\x_{t-1})\right)^T \EE\left[\left(\g(\x_t,\zb_t) - \gb(\x_t)\right)\rvert\cF_t\right] \nonumber\\ 
	& \leq 	(1-\beta_t)^2\norm{\y_t-\gb(\x_{t-1})}^2  +	\beta_t^2V_g. \label{track1a}
	\end{align}
	where the last inequality follows from \eqref{ainner}. Taking full expectation in \eqref{track1a} and plugging into \eqref{pep}, we obtain the required result as
	\begin{align}
	&\E{\norm{\y_{t+1} -\gb(\x_t) }^2 } \\
	&\leq (1+\beta_t)(1-\beta_t)^2\E{\norm{\y_t-\gb(\x_{t-1})}^2}+(1+\beta_t)\beta_t^2V_g+ \frac{1-\beta_t^2}{\beta_t}C_g\E{\norm{ \x_t-\x_{t-1} }^2} \\
	& \leq (1-\beta_t)\E{\norm{\y_t-\gb(\x_{t-1})}^2} + 2V_g\beta_t^2 + 2\frac{\alpha_t^2}{\beta_t} C_fC_g^2 + 2\frac{\delta_t^2}{\beta_t} JC_{\ell}^2 C_g C_q C_h.
	\end{align} 
	where we have also used the fact that $\beta_t < 1$. To establish the upper bound, take $D_y := \E{\norm{\y_2 - \gb(\x_1)}^2} + 2V_g + 2C_g\br{ C_fC_g + JC_{\ell}^2 C_q C_h}$ and observe that $\E{\norm{\y_{3} -\gb(\x_2)}^2} \leq (1-\beta_t)D_y+2V_g\beta_t^2 + 2\frac{\alpha_t^2}{\beta_t} C_fC_g^2 + 2\frac{\delta_t^2}{\beta_t} JC_{\ell}^2 C_g C_q C_h \leq D_y$ since $\alpha_t/\beta_t \leq 1$ and $\delta_t/\beta_t \leq 1$. The argument can inductively be applied to all values of $t$. 
	
	The derivation of \eqref{htrack} follows along similar lines. For the upper bound, we set  \\$D_z = \E{\norm{\z_2 - \hb(\x_1)}^2} + 2V_h + 2C_h\br{ C_fC_g + JC_{\ell}^2 C_q C_h}$ and proceed inductively as before. 
\end{IEEEproof}

Next, we analyze the distance from the optimal set $\norm{ \x_t -\x^{\gamma} }^2$, and establish its relation with the optimality gap and the constraint violation. 
	\begin{lemma}\label{optG}
For convex functions $F$ and $\Q$, we have the following bound. 
		\begin{align}
		\E{\norm{\x_{t+1}-\x^{\gamma}}^2} &\leq  \E{\norm{\x_t-\x^{\gamma}}^2} - 2\alpha_t \E{F(\x_t) + \frac{\delta_t}{\alpha_t} L(\x_t) - F(\x^{\gamma})} + 2\alpha_t^2 C_fC_g  \nonumber\\
		&\hspace{5mm}  + 2\delta_t^2 J C_{\ell}^2 C_q C_h + \beta_t \E{\norm{\y_{t+1} - \gb(\x_t)}^2} + \beta_t \E{\norm{\z_{t+1} - \hb(\x_t)}^2}  \nonumber\\
		&\hspace{5mm} + \frac{\alpha_t^2}{\beta_t}L_fD_x C_g + \frac{\delta_t^2}{\beta_t}\br{\sqrt{J}L_qC_{\ell} + C_q}D_x C_h. \label{optGap} 
		\end{align} 
	\end{lemma}
\begin{IEEEproof}
Since $\x^\gamma \in \cX$, the non-expansiveness of the $\Pi_{\cX}$ operator implies that 
	\begin{align}
	\norm{\x_{t+1}-\x^{\gamma}}^2 &= \norm{\Pi_{\cX}\bc{\x_t - \alpha_t\sdF\g(\x_t,\zb_t)\dF f(\y_{t+1}) - \delta_t\sdF \h(\x_t,\zb_t)\dF p(\z_{t+1})} - \x^{\gamma}}^2 \\
	& \leq \norm{\x_t - \x^{\gamma} -\br{\alpha_t\sdF\g(\x_t,\zb_t)\dF f(\y_{t+1}) + \delta_t\sdF \h(\x_t,\zb_t)\dF p(\z_{t+1})}}^2 \\
	& = \norm{\x_t-\x^{\gamma}}^2 + \norm{\alpha_t\sdF\g(\x_t,\zb_t)\dF f(\y_{t+1}) + \delta_t\sdF \h(\x_t,\zb_t)\dF p(\z_{t+1})}^2 \\
	&\hspace{5mm} - 2\br{\x_t - \x^{\gamma}}^T\br{\alpha_t\sdF\g(\x_t,\zb_t)\dF f(\y_{t+1}) + \delta_t\sdF \h(\x_t,\zb_t)\dF p(\z_{t+1})} \\
	& = \norm{\x_t-\x^{\gamma}}^2 + \norm{\x_{t+1}-\x_t}^2 + u_t + v_t\\
	&\hspace{5mm}- 2\br{\x_t - \x^{\gamma}}^T\br{\alpha_t\sdF\g(\x_t,\zb_t)\dF f(\gb(\x_t)) + \delta_t\sdF \h(\x_t,\zb_t)\dF p(\hb(\x_t))}\label{optG2}
	\end{align}
	where $u_t = 2\alpha_t\br{\x_t - \x^{\gamma}}^T\sdF\g(\x_t,\zb_t) \br{\dF f(\gb(\x_t)) - \dF f(\y_{t+1})}$, and \\$v_t = 2\delta_t\br{\x_t - \x^{\gamma}}^T\sdF\h(\x_t,\zb_t) \br{\dF p(\hb(\x_t)) -\dF p(\z_{t+1})}$. 
	Taking conditional expectation given $\cF_t$ in \eqref{optG2}, we obtain
	\begin{align}
	\E{\norm{\x_{t+1}-\x^{\gamma}}^2\lvert\cF_t} &\leq  \norm{\x_t-\x^{\gamma}}^2 + \E{\norm{\x_{t+1}-\x_t}^2\lvert\cF_t} + \E{u_t\lvert\cF_t} + \E{v_t\lvert\cF_t} \nonumber\\
	&\hspace{2.5mm} - 2\br{\x_t - \x^{\gamma}}^T \E{\alpha_t\sdF\g(\x_t,\zb_t)\dF f(\gb(\x_t)) + \delta_t\sdF \h(\x_t,\zb_t)\dF p(\hb(\x_t))\lvert\cF_t} \nonumber\\
	& = \norm{\x_t-\x^{\gamma}}^2 + 2\alpha_t^2 C_fC_g + 2\delta_t^2 J C_{\ell}^2 C_q C_h + \E{u_t\lvert\cF_t} + \E{v_t\lvert\cF_t} \nonumber\\
	&\hspace{2.5mm} - 2\br{\x_t - \x^{\gamma}}^T\br{\alpha_t\dF F(\x_t) + \delta_t\dF L(\x_t)}\\
	& \leq \norm{\x_t-\x^{\gamma}}^2 + 2\alpha_t^2 C_fC_g + 2\delta_t^2 J C_{\ell}^2 C_q C_h + \E{u_t\lvert\cF_t} + \E{v_t\lvert\cF_t} \nonumber\\
	&\hspace{2.5mm} - 2\alpha_t \br{F(\x_t) - F(\x^{\gamma})} - 2\delta_t (L(\x_t)-L(\x^\gamma))\label{optG3}.
	\end{align}
	where \eqref{optG3} follows from Lemma \ref{prilG} and the convexity of functions $F$  and $L$. Next, noting that $L(\x^\gamma) = 0$ and taking full expectation in \eqref{optG3}, we obtain
	\begin{align}
	\E{\norm{\x_{t+1}-\x^{\gamma}}^2} &\leq  \E{\norm{\x_t-\x^{\gamma}}^2} - 2\alpha_t \br{F(\x_t) - F(\x^{\gamma})} - 2\delta_t L(\x_t) + 2\alpha_t^2 C_fC_g \nonumber\\
	&\hspace{5mm} + 2\delta_t^2 J C_{\ell}^2 C_q C_h  + \E{u_t} + \E{v_t} \label{intm}.  	
	\end{align}
	In order to bound $u_t$, we utilize the Cauchy-Schwartz and Peter-Paul inequalities as 
	\begin{align}
	u_t &= 2\alpha_t\br{\x_t-\x^{\gamma}}^T\sdF\g(\x_t,\zb_t) \br{\dF f(\gb(\x_t)) - \dF f(\y_{t+1})}\\
	& \leq 2\alpha_t L_f\norm{\x_t-\x} \norm{\y_{t+1} - \gb(\x_t)} \norm{\sdF\g(\x_t,\zb_t)} \\
	& \leq \beta_t  \norm{\y_{t+1} - \gb(\x_t)}^2 + \frac{\alpha_t^2}{\beta_t}L_f\norm{\x_t-\x}^2 \norm{\sdF \g(\x_t,\zb_t)}^2 \\
	& \leq \beta_t  \norm{\y_{t+1} - \gb(\x_t)}^2 + \frac{\alpha_t^2}{\beta_t}L_fD_x \norm{\sdF \g(\x_t,\zb_t)}^2. \label{optG4}
	\end{align}
	where \eqref{optG4} follows from \eqref{aslater}. Taking expectation on both sides of \eqref{optG4}, we obtain $E{u_t} \leq \beta_t \E{\norm{\y_{t+1} - \gb(\x_t)}^2} + \frac{\alpha_t^2}{\beta_t}L_fD_x C_g$,
%	\begin{align}
%	\E{u_t} &\leq \beta_t \E{\norm{\y_{t+1} - \gb(\x_t)}^2} + \frac{\alpha_t^2}{\beta_t}L_fD_x C_g.\label{utbound}
%	\end{align}
	and we can also obtain $\E{v_t} \leq \beta_t \E{\norm{\z_{t+1} - \hb(\x_t)}^2} + \frac{\delta_t^2}{\beta_t}\br{\sqrt{J}L_qC_{\ell} + C_q}D_x C_h$.
%	\begin{align}
%	\E{v_t} \leq \beta_t \E{\norm{\z_{t+1} - \hb(\x_t)}^2} + \frac{\delta_t^2}{\beta_t}\br{\sqrt{J}L_qC_{\ell} + C_q}D_x C_h. \label{vtbound}
%	\end{align}
	The required result follows from substituting into \eqref{intm}. 
\end{IEEEproof}

Having established the required recursive relationships, we proceed with applying telescopic sums in order to obtain the required bounds. Let
\begin{align}
I_t &:= \E{\norm{\x_t - \x^{\gamma}}^2 + \norm{\y_t - \gb(\x_{t-1})}^2 +  \norm{\z_t - \hb(\x_{t-1})}^2} 
\end{align}
and observe that $I_t \leq D_x + D_y + D_z =: D_1$. Multiplying \eqref{gtrack} and \eqref{htrack} with $\br{1+\beta_t}$ and substituting in \eqref{optGap},  we obtain
\begin{align}
I_{t+1} &\leq I_t - 2\alpha_t \E{F(\x_t) + \frac{\delta_t}{\alpha_t} L(\x_t) - F(\x^{\gamma})}  + 2\delta_t^2 \br{C_fC_g + J C_{\ell}^2 C_q C_h} + 4\br{V_g + V_h}\beta_t^2 \nonumber \\
&\hspace{2.5mm}+ \frac{\delta_t^2}{\beta_t}\br{4\br{C_g + C_h}C_fC_g + 4\br{C_g + C_h} J C_{\ell}^2 C_q C_h +  \br{ L_fC_g+ \sqrt{J}L_qC_hC_{\ell} + C_hC_q}D_x} \nonumber\\
&\leq I_t - 2\alpha_t \E{F(\x_t) + \frac{\delta_t}{\alpha_t} L(\x_t) - F(\x^{\gamma})} + D_2\br{\delta_t^2 + \beta_t^2 + \frac{\delta_t^2}{\beta_t} }.
\end{align}
where $D_2$ is the maximum of $4\br{C_g + C_h}C_fC_g + 4\br{C_g + C_h} JC_{\ell}^2 C_q C_h \\+  \br{ L_fC_g+ \sqrt{J}L_qC_hC_{\ell} +C_hC_q}D_x$, $2\br{C_fC_g + C_{\ell}^2 C_q C_h}$, and $4\br{V_g + V_h}$. Summing over \\$t = T/2$ to $T$, noting that $\alpha_t \leq \alpha_{T/2}$, and rearranging, we obtain the required result as
\begin{align}
	\sum_{t = T/2}^T &\E{F(\x_t) +\frac{\delta_t}{\alpha_t} L(\x_t) - F(\x^{\gamma})}\\
	& \leq \frac{{I_{T/2}}}{\alpha_{T/2}} + \sum_{t = T/2}^T{I_t}\br{\frac{1}{\alpha_t} - \frac{1}{\alpha_{t-1}}} - \frac{{I_{T+1}}}{\alpha_{T+1}} + D_2\br{\sum_{t = T/2}^T\bs{\frac{\delta_t^2}{\alpha_t} + \frac{\beta_t^2}{\alpha_t} + \frac{\delta_t^2}{\alpha_t\beta_t} } } \\
	& \leq \frac{D_1}{\alpha_{T/2}} + D_1\sum_{t = T/2 + 1}^T\br{\frac{1}{\alpha_t} - \frac{1}{\alpha_{t-1}}} + D_2\br{\sum_{t = T/2}^T\bs{\frac{\delta_t^2}{\alpha_t} + \frac{\beta_t^2}{\alpha_t} + \frac{\delta_t^2}{\alpha_t\beta_t} } } \\
	& \leq \frac{D_1}{\alpha_T} + D_2\br{\sum_{t = T/2}^T\bs{\frac{\delta_t^2}{\alpha_t} + \frac{\beta_t^2}{\alpha_t} + \frac{\delta_t^2}{\alpha_t\beta_t} } }  \label{omegaexp}
\end{align}
where we have dropped the negative terms from the right-hand side. 
which is the required result. 

\section{Proof of Lemma \ref{perturb}}\label{pertproof}
Let us associate dual variable $\lam$ with the constraint in \eqref{gammaprob} and let $\lam^\gamma \geq 0$ be the dual optimal solution. Since $\gamma < \sigma/2 < \sigma$, there exists a strictly feasible $\xt$ such that $Q_j(\xt)+\gamma < Q_j(\xt) + \sigma \leq 0$, and consequently strong duality holds for \eqref{gammaprob}. Therefore, we have that
	\begin{align}
	F(\x^{\gamma}) = \min_{\x \in\cX} F(\x) + \ip{\lam^{\gamma}}{\Q(\x) + \gamma\vone}   \leq F(\x^{\star}) + \ip{\lam^{\gamma}}{\Q(\x^\star) + \gamma\vone} \leq F^\star +\gamma\vone^T \lam^{\gamma}, \label{pert2}
	\end{align}
	where we have used the fact that $\Q(\x^\star) \leq 0$. Let $\xt$ be a strictly feasible point specified in \eqref{aslater}. Then from \eqref{pert2}, it also follows that
	\begin{align}
F(\x^{\gamma}) & \leq F(\xt) + \ip{\lam^{\gamma}}{\Q(\xt) + \gamma\vone} \leq F(\xt) + (\gamma-\sigma)\vone^T\lam^\gamma
\end{align}	
or equivalently
\begin{align}
\vone^T\lam^\gamma \leq \frac{F(\xt)-F(\x^\gamma)}{\sigma-\gamma}\leq \frac{2\sqrt{C_gC_f}D_x}{\sigma} \label{sumlam}
\end{align}
where the last inequality follows since $F$ is Lipschitz continuous with parameter $\sqrt{C_gC_g}$ and since $\cX$ is compact. The required bound follows from substituting \eqref{sumlam} into \eqref{pert2}. 

\section{Numerical verification of convexity of the formulation in Example 2} \label{numver}
\begin{lemma}
	The following function is convex with respect to $\lambda$, and $p$ for the box constraints $0.1\leq \lambda \leq 15$, and $14\leq p \leq 100$ 
	\begin{align}\label{conv}
	U(\lambda,p) =  \frac{\lambda \EE\left[\br{\frac{1}{B\log(1+p\zeta)}}^2\right]}{2\left(1-\lambda\EE\left[\frac{1}{B\log(1+p\zeta)}\right]\right)} - 0.1\log(\lambda), 
	\end{align}
	where $\zeta$ is a Chi-squared distributed random variable with $2K$ degrees of freedom. 
\end{lemma}

\begin{IEEEproof}
	Since the expression in \eqref{conv} is of non linear expectation function, it is more difficult to get an analytical proof. Instead, we try to derive the Hessian of \eqref{conv} numerically, and show that the minimum eigenvalue of Hessian is nonnegative.  To do that, we take help of two MATLAB functions namely `numerical gradient', and `numerical integration'. The results in Fig. \ref{conPf} confirm that the Hessian of the expression in \eqref{conv} is positive semidefinite matrix.
	
	\begin{figure*}
		\centering
		\includegraphics[width=\linewidth, height = 0.5\linewidth]
		{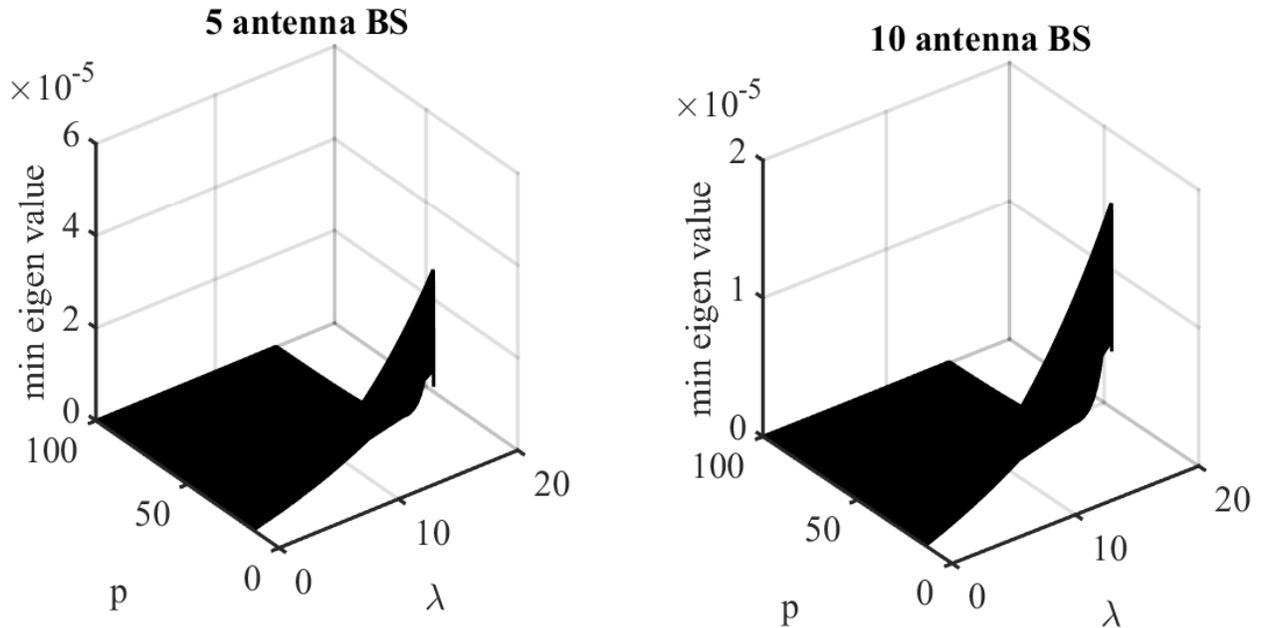}
		\caption{minimum eigenvalue of the Hessian of \eqref{conv} with K = 5, and K = 10 antennas at the BS.}
		\label{conPf}
	\end{figure*}
	
\end{IEEEproof}  	 
 \footnotesize
    \bibliographystyle{IEEEtran}
    \bibliography{IEEEabrv,references}

\end{document}